\documentclass[12pt,preprint]{aastex}
\usepackage{epsfig}
\usepackage{subfigure}
\usepackage{amssymb}

\renewenvironment{figure}{\begin{figure*} }{\end{figure*}}

\newcommand{\bu}{{\bf u}}

\newcommand{\bg}{{\bf g}}

\newcommand{\be}{{\bf e}}
\newcommand{\grad}{{\mathbf \nabla}}
\renewcommand{\div}{{\mathbf \nabla} \cdot}

\newcommand{\ephi}{\be_\phi}
\newcommand{\er}{\be_r}
\newcommand{\ex}{\be_x}

\newcommand{\ez}{\be_z}
\newcommand{\dd}{{\rm d}}
\newcommand{\bO}{\mbox{\boldmath $\Omega$}}
\newcommand{\Enu}{{\rm E}_\nu}

\begin{document}

\title{On the Penetration of Meridional Circulation below the Solar Convection Zone II: Models with Convection Zone, the Taylor-Proudman constraint and Applications to Other Stars.}

\author{P. Garaud \& L. Acevedo-Arreguin} 

\affil{Department of Applied Mathematics and Statistics, Baskin School of Engineering, University of California Santa Cruz, 1156 High Street, CA 95064 
Santa Cruz, USA}

\maketitle

\section*{Abstract}

The solar convection zone exhibits a strong level of 
differential rotation, whereby
the rotation period of the polar regions is about 25-30\% longer than 
the equatorial regions. The Coriolis force associated with these zonal
flows perpetually ``pumps'' the convection zone fluid, and maintains 
a quasi-steady circulation, poleward near the surface.
What is the influence of this meridional circulation on the underlying 
radiative zone, and in particular, does it provide a significant source of 
mixing between the two regions? In Paper I, we began to study this question
by assuming a fixed meridional flow pattern in the convection zone and 
calculating its penetration depth into the radiative zone. We found that
the amount of mixing caused depends very sensitively on the assumed 
flow structure near the radiative--convective interface. We continue
this study here by including a simple model for the convection zone ``pump'', 
and calculating in a self-consistent manner the 
meridional flows generated in the whole Sun. We find that the global 
circulation timescale depends in a crucial way on two factors: the overall 
stratification of the radiative zone as measured by the Rossby number times 
the square root of the Prandtl number, and, for weakly stratified systems, 
the presence or absence of stresses within the radiative zone capable of
breaking the Taylor-Proudman constraint. We conclude by discussing the 
consequences of our findings for the solar interior and argue that a 
potentially important mechanism for mixing in Main Sequence stars has so far
been neglected. 



\section{Introduction}


Various related mechanisms are thought to contribute to the 
generation and maintenance of large-scale meridional flows in the 
solar convection zone. The effect of rotation on turbulent convection 
induces a relatively 
strong anisotropy in the Reynolds stresses (Kippenhahn 1963), 
in particular near the base
of the convection zone where the convective turnover time is of the order
of the solar rotation period. The divergence of these anisotropic stresses
can directly drive large-scale meridional flows (see R\"udiger, 1989, for a 
discussion of this effect). It also drives 
large-scale zonal flows (more commonly referred 
to as differential rotation) which then induce meridional forcing 
through the biais of the Coriolis force, a mechanism referred to 
as ``gyroscopic pumping'' (McIntyre 2007). 
Indeed, the polar regions of the convection zone are observed to be 
more slowly rotating than the bulk of the Sun (Schou {\it et al.} 1998), 
so that the associated Coriolis force in these regions drives fluid 
{\it towards} the polar axis. Meanwhile, equatorial 
regions are rotating more rapidly than the average, and are therefore 
subject to a Coriolis force pushing fluid {\it away} 
from the polar axis. The most
likely flow pattern resulting from the combination of these forces is 
one with an equatorial upwelling, a surface poleward flow
and a deep return flow. This pattern is indeed observed
near the solar surface: poleward surface and sub-surface flows 
with velocities up to a few tens of meters per second have been observed by 
measurements of photospheric line-shifts (Labonte \& Howard 1982) and
by time-distance helioseismology (Giles et al. 1997) respectively. 

The amplitude and spatial distribution of these meridional flows deeper in 
the convection zone remains essentially unknown, as the sensitivity of 
helioseismic methods rapidly drops below the surface. As a result, the 
question of whether some of the pumped mass flux actually penetrates
into the underlying radiative zone is still open, despite its obvious
importance for mixing of chemical species (Pinsonneault, 1997; 
Elliott \& Gough, 1999), and its presumed role in the dynamical 
balance of the solar interior (Gough \& McIntyre 1998, McIntyre, 2007,
Garaud, 2007, Garaud \& Garaud, 2008) and in some models of the solar
dynamo (see Charbonneau, 2005 for a review). 

In Paper I (Garaud \& Brummell, 2008),
we began a systematic study of the penetration of 
meridional flows from the convection zone into the radiative zone
by considering a related but easier question: {\it assuming that the amplitude
and geometry of meridional flows in the convection zone are both known}, 
what is their influence on the underlying radiative zone? This
simpler question enabled us to study the dynamics of the radiative zone
only by assuming a flow profile at the radiative--convective interface 
(instead of having to include the more complex convection zone in the 
calculation). The overwhelming conclusion of that first study was that 
the degree to which flows penetrate into the stratified interior (in the model) 
is {\it very} sensitive to the interfacial conditions selected. Hence, 
great care must be taken when using a ``radiative-zone-only'' 
model to make definite predictions about interior flow amplitudes. 
In addition, that approach makes the implicit assumption 
that the dynamics of the radiative zone do not in return influence 
those of the convection zone, but the only way to verify this 
is to construct a model which includes both regions. 
This was the original purpose of the present study; as we shall see, 
the combined radiative--convective model we construct here provides 
insight into a much broader class of problems. 

We therefore propose a simplified model of 
the Sun which includes both a ``convective'' region and a ``radiative''
region, where the convective region is forced in such a way as to 
promote gyroscopic pumping of meridional flows. We calculate the 
flow solution everywhere and characterize how it scales in terms of 
governing parameters
(e.g. stellar rotation rate, stratification, diffusivities, etc..), focusing
in particular on the flows which are entering the radiative zone. 
We begin with a simple Boussinesq Cartesian model 
(Section \ref{sec:cart}), 
first in the unstratified limit (Section \ref{sec:unstrat}) 
and then in the more realistic case of a radiative--convective 
stratification (Section \ref{sec:strat}). Although the Cartesian
results essentially illustrate most of the relevant physical phenomena, 
we confirm our analysis with numerical
solutions of the full set of equations 
in a spherical geometry in Section \ref{sec:spherical}. We then use
this information in Section \ref{sec:disc} 
to discuss the effects of mixing by meridional flows 
both in the Sun and in other Main Sequence stars. 

\section{A Cartesian model}
\label{sec:cart}

\subsection{Model setup}

As in Paper I, we first study the problem in a Cartesian geometry.
Since our primary aim is to understand the behavior of the meridional 
flows generated (e.g. scaling of the solutions) 
in terms of the governing parameters, this approach is sufficient and
vastly simplifies the required algebra. In Section \ref{sec:spherical}, 
we turn to numerical simulations to study the problem in a spherical geometry.

In this Cartesian model section, distances 
are normalized to the solar radius $R_\odot$, and velocities to  
$R_\odot \Omega_\odot$ where $\Omega_\odot$ is the mean solar angular 
velocity (the exact value is not particularly relevant here).
The coordinate system is $(x,y,z)$, where
$x$ should be thought of as the azimuthal coordinate $\phi$, with 
$x \in [0,2\pi]$; $y$ represents minus the co-latitude and
spans the interval  $y \in [0,\pi]$ (the poles are at
$y=0$ and $y = \pi$ while the equator is at $y = \pi/2$). 
Finally the $z$-direction is the radial direction with 
$z \in [0, 1] $, and represents the direction of (minus) gravity so that 
$z=0$ is the interior, and $z=1$ is the surface.  

In this framework, the system rotates with mean angular 
velocity $\bO = (0,0,1)$,
thereby implicitly assuming that the rotation axis is everywhere aligned with 
gravity. This assumption induces another ``geometric'' error in the velocity
estimates for the meridional flows, comparable with the error made in reducing 
the problem to a Cartesian analysis; it does not influence the predicted
scalings (except in small equatorial regions which we ignore here). 

\begin{figure}[h]
\centerline{\epsfig{file=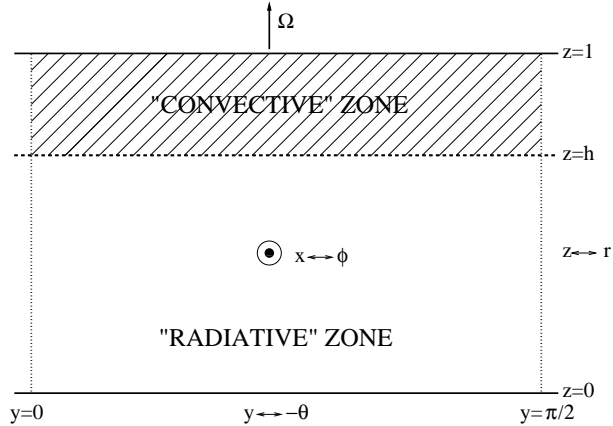,width=8cm}}
\caption{Cartesian model geometry and intended correspondence with the 
spherical case. The shaded area marks the convective region, where forcing 
is applied. The $y=0$ and $y=\pi/2$ lines mark the ``poles'' 
and the ``equator''. The system is assumed to be periodic with period 
$\pi$ in the $y-$direction. 
}
\label{fig:model} 
\end{figure}

We divide the domain in two regions, by introducing the dimensionless
constant $h$ to represent the radiative--convective interface. 
Thus $z \in [0,h]$ represents the ``radiative zone'' 
while $z \in [h,1]$ represents the ``convection zone''. 
From here on, $h = 0.7$. Figure \ref{fig:model} illustrates 
the geometry of the Cartesian system.

\subsection{Model equations}
\label{sec:modeleq}

For this simple Cartesian approach, we work with the Boussinesq approximation
(this assumption is dropped in the spherical model but doesn't affect the 
predicted scalings). The background state is assumed to be stratified, 
steady, and in hydrostatic equilibrium. The background density and 
temperature profiles are denoted by $\bar T (z)$ and $\bar \rho(z)$ 
respectively. Density and temperature 
perturbations to this background state ($\rho$ and $T$) 
are then assumed to be linearly related: $\rho = -  \alpha_T(z) T$, 
where $\alpha_T(z)$ is the coefficient of thermal expansion. 
For simplicity, we will assume that the background temperature gradient 
$\bar T_z$ is constant throughout
the domain $z\in[0,1]$, and treat the convection zone as a region 
where $\alpha_T \rightarrow 0$ while the radiative zone has 
$\alpha_T \neq 0$ (see below). The alternative option of using a 
constant $\alpha_T$ and a varying $\bar T_z$ yields 
qualitatively equivalent scalings for the meridional flows
(a statement which is verified in Section 
\ref{sec:spherical}) although the algebra is trickier. 

The set of equations governing the system in this approximation  
are the momentum, mass and thermal energy conservation 
equations respectively:
\begin{eqnarray}
\label{eq:gov_eqs}
\frac{\partial \bu}{\partial t} + \bu \cdot \grad \bu +  2 \ez \times \bu = - \nabla p + {\rm Ro}^2(z) T  \ez + \Enu \grad^2 \bu \mbox{   ,  } \nonumber \\
\div\bu = 0 \mbox{   ,  } \nonumber \\
\frac{\partial T}{\partial t} + \bu \cdot \grad T + \bu \cdot \ez = \frac{\Enu}{\rm Pr} \grad^2 T \mbox{   ,  } 
\end{eqnarray}
where $\bu = (u,v,z)$ is the velocity field. In these equations, 
temperature perturbations have been normalized to the background
temperature difference across the box $R_\odot \bar T_z$. The governing 
non-dimensional parameters are:
\begin{eqnarray}
{\rm Ro}(z)  = N(z)/\Omega_\odot \mbox{ the Rossby number }  \mbox{   ,  } \nonumber \\
\Enu = \frac{\nu}{R_\odot^2 \Omega_\odot} \mbox{ the Ekman number } \mbox{   ,  }  \nonumber \\
{\rm Pr} = \frac{\nu}{\kappa_T} \mbox{ the Prandtl number } \mbox{   ,  } 
\end{eqnarray}
where the dimensional quantity $N^2(z) = \alpha_T(z) \bar T_z g$ 
is the square of the 
Brunt-V\"ais\"al\"a frequency ($g$ is the magnitude of gravity and is assumed to be constant). The microscopic diffusion coefficients $\nu$ (the viscosity) 
and $\kappa_T$ (the thermal conductivity) are both assumed to be constant. 
In the Sun, near the radiative--convective interface, $\Enu \simeq 2\times 
10^{-15}$ and Pr $\simeq 2\times 10^{-6}$ (Gough, 2007). 

As mentioned earlier, the transition between the model radiative zone and 
convection zone is measured by the behavior of $\alpha_T(z)$, 
which goes from 0 for $z > h$ to a finite value for $z<h$. This 
can be modeled in a non-dimensional way through the Rossby number, which 
we assume is of the form: 
\begin{equation}
{\rm Ro}(z) = \frac{{\rm Ro}_{\rm rz}}{2} \left[1+\tanh\left(\frac{h-z}{\Delta}\right)\right] \mbox{   ,  } 
\end{equation}
where ${\rm Ro}_{\rm rz}$ is constant. The lengthscale $\Delta$ may be 
thought of as the thickness of the ``overshoot'' region 
near the base of the convection zone, but in practise is mostly used to
ensure continuity and smoothness of the background state through the tanh 
function.

In what follows, we further restrict our study to an ``axially'' symmetric 
$(\partial/\partial x = 0)$, steady-state $(\partial/\partial t = 0)$ problem.
Within the radiative zone, the nonlinear terms $\bu \cdot \grad \bu$ 
and $\bu \cdot \grad T$ are assumed to be negligible. 
Within the convection zone on the other hand, anisotropic 
turbulent stresses are thought to drive the observed differential
rotation\footnote{Note that for slowly rotating stars, such as the Sun, 
the direct generation of meridional flows by anisotropic stresses 
is a much weaker effect in the bulk of the convection zone. We neglect it here.}. We model this effect in the simplest possible way, replacing
the divergence of the stresses by a linear relaxation towards the 
observed convection zone profile:
\begin{equation}
\bu \cdot \grad \bu \rightarrow {\bf F}_{\rm turb} = \frac{ \bu - \bu_{\rm cz}}{\tau} \mbox{   ,  } 
\end{equation}
where $\bu_{\rm cz} = u_{\rm cz}(y,z) \ex$ and the function 
$u_{\rm cz}(y,z)$ models the observed azimuthal velocity profile in the solar
convection zone. This is analogous to the prescription used 
by Spiegel \& Bretherton (1968) in their study of the effect of a 
convection zone on solar spin-down, although in their model the 
convection zone was not differentially rotating.
The dimensionless relaxation timescale $\tau$ can be thought of, 
for example, as 
being of the same order of magnitude as the convective turnover time divided
by the rotation period. It is modeled as:
\begin{equation}
\label{eq:tau_profile}
\tau^{-1}(z) = \frac{\Lambda}{2} \left[1+\tanh\left(\frac{z-h}{\Delta}\right)\right]\mbox{   .  } 
\end{equation}
Note that in the real solar convection zone, $\tau$ varies 
by orders of magnitudes between the surface ($\tau \sim 10^{-3}$) 
and the bottom of the convection zone ($\tau \sim 1$).  Here, we assume
that $\Lambda$ is constant for simplicity.

We adopt the following profile for $u_{\rm cz}(y,z)$: 
\begin{eqnarray}
&& u_{\rm cz}(y,z) = \frac{U_0(z)}{2} \left[1+\tanh\left(\frac{z-h}{\Delta}\right)\right] e^{iky} = \hat u_{\rm cz}(z) e^{iky} \mbox{   ,  } \nonumber \\
&& U_0(z) = U_0(h) + S(z-h)\mbox{   ,  } 
\end{eqnarray}
where $k = 2$ to match the equatorial symmetry of the 
observed solar rotation profile. The tanh function once again is merely 
added to guarantee continuity of the forcing across the overshoot layer. 
The function $U_0(z)$ describes the imposed ``vertical shear'', and is 
for simplicity taken to be a linear function of $z$.
If $U_0(h) = 0$, the forcing effectively vanishes at the base 
of the convection zone. If $U_0(h) \neq 0$ on the other hand, a strong 
azimuthal shear is forced at the interface. The observed solar 
rotation profile 
appears to be consistent with $U_0(h)$ and $S$ both being non-zero 
(and of the 
order of 0.1, although since we are studying a linear problem, the amplitude
of the forcing is somewhat irrelevant). Note that if $S = 0$ the forcing 
velocity $\bu_{\rm cz}$ has zero vorticity.         

Finally, the observed asphericity in the temperature profile 
is negligible in the solar convection zone; this is attributed to 
the fact that 
the turbulent convection very efficiently mixes heat both vertically 
and horizontally. We model this effect as:
\begin{equation}
\bu \cdot \grad T \rightarrow - D(z) \grad^2 T \mbox{   ,  } 
\end{equation}
where the turbulent heat diffusion coefficient is modeled as 
\begin{equation}
\label{eq:D_profile}
D(z) = \frac{D_0}{2} \left[1+\tanh\left(\frac{z-h}{\Delta}\right)\right]\mbox{   ,  } 
\end{equation}
and thus vanishes beneath the overshoot layer. We will assume that the 
diffusion timescale $1/D_0$ (in non-dimensional units) is much smaller 
than any other typical timescale in the system ($D_0 \gg 1$). 

Projecting the remaining equations into the Cartesian coordinate system, 
and seeking solutions in the form $q(y,z) = \hat q(z) e^{iky}$ for each of the unknown
quantities yields
\begin{eqnarray}
&&-2 \hat v = \Enu \left(\frac{\dd^2 \hat u}{\dd z^2} - k^2 \hat u \right) - \frac{\hat u-\hat u_{\rm cz}}{\tau} \mbox{   ,   }  \nonumber \\
&&2 \hat u = - ik \hat p + \Enu \left( \frac{\dd^2 \hat v}{\dd z^2} - k^2 \hat v \right) - \frac{\hat v}{\tau} \mbox{   ,   }  \nonumber \\
&&0 = - \frac{\dd \hat p}{\dd z} + {\rm Ro}^2(z) \hat T + \Enu \left( \frac{\dd^2 \hat w}{\dd z^2} - k^2 \hat w \right) - \frac{\hat w}{\tau}  \mbox{   ,   }  \nonumber \\
&& ik \hat v  + \frac{\dd \hat w}{\dd z} = 0 \mbox{   ,   } \nonumber \\
&& \hat w = \left(\frac{\Enu}{\rm Pr} + D(z) \right) \left( \frac{\dd^2 \hat T}{\dd z^2} - k^2 \hat T \right)\mbox{   ,   } 
\label{eq:maineqs}
\end{eqnarray}
Note that as required, the imposed forcing term drags the fluid in 
the azimuthal direction: for $\tau\rightarrow 0$, 
$\hat u \rightarrow \hat u_{\rm cz}$ in the convection zone. 
The meridional flows $\hat v$ and 
$\hat w$ on the other hand are generated by the $y-$component of the 
Coriolis force and by mass conservation respectively (the essence of 
gyroscopic pumping, see McIntyre 2007). 

We now proceed to solve these equations to gain a better
understanding of the meridional flows and their 
degree of penetration into the radiative zone below.
We use a dual approach, solving these equations first analytically 
under various limits, and then exactly using a simple 
Newton-Raphson-Kantorovich (NRK) two-point boundary value algorithm. 
The analytical approximations yield predictions for the relevant 
scalings of the solutions in terms of the governing parameters (an in 
particular, the Ekman number and the Rossby number) 
which are then confirmed by the exact numerical solutions. 

\subsection{The unstratified case}
\label{sec:unstrat}

Although this limit is not a priori relevant to the physics of the 
solar interior, we begin by studying the case of an unstratified region, 
setting ${\rm Ro}_{\rm rz} = 0$ (in this case, the thermal energy equation can
be discarded). This simpler problem, as we shall demonstrate, contains the 
essence of the problem.

In order to find analytical approximations to the solutions, 
we solve the governing equations separately in the convective zone
and in the radiative zone. At this point, it may be worth pointing out that 
in the unstratified case, the nomenclatures ``convective'' and ``radiative'' 
merely refer to regions which respectively 
are and are not subject to the additional forcing. 

We assume that the transition region is very 
thin\footnote{More precisely, $\Delta \ll \Enu^{1/2}$, see Section \ref{sec:twist}.}. In this case, $\tau^{-1} = \Lambda$ 
for $z > h$ while $\tau^{-1} = 0 $ for $z < h$.  Similarly, 
$\hat u_{\rm cz}(z) = U_0(h) + S(z-h)$ in the convection zone 
while $\hat u_{\rm cz}(z) =  0$ in the radiative zone. Once obtained, 
the solutions are patched at the radiative--convective interface.

\subsubsection{Solution in the convection zone}
\label{sec:unstr_conv_ss}

In the convection zone, the equations reduce to 
\begin{eqnarray}
&&-2 \hat v = - \Lambda(\hat u-\hat u_{\rm cz}) \mbox{   ,   }  \nonumber \\
&&2 \hat u = - ik \hat p  - \Lambda \hat v  \mbox{   ,   }  \nonumber \\
&&0 = - \frac{\dd \hat p}{\dd z} - \Lambda \hat w  \mbox{   ,   }  \nonumber \\
&& ik \hat v  + \frac{\dd \hat w}{\dd z} = 0 \mbox{   ,   } 
\label{eq:cartconv}
\end{eqnarray}
where we have neglected the viscous dissipation terms in favor of the 
forcing terms since $E_\nu \ll \Lambda$ for all reasonable solar parameters. 
Combining them yields 
\begin{equation}
\label{eq:unstr_conv_weq}
\frac{\dd^2 \hat w}{\dd z^2} = \frac{k^2 \Lambda^2}{4+\Lambda^2} \hat w + 2ik \frac{\Lambda}{4+\Lambda^2} \frac{\dd  \hat u_{\rm cz}}{\dd z} \mbox{   .   } 
\end{equation}
This second-order ordinary differential equation\footnote{The original order of the system is much reduced in the convection zone since we ignored the effect of viscous terms there.} for $\hat w (z)$ 
suggests the introduction of a new lengthscale 
\begin{equation}
\delta = \frac{\sqrt{4 + \Lambda^2}}{k \Lambda} \mbox{   ,   } 
\end{equation}
so that the general solution to (\ref{eq:cartconv}) is
\begin{eqnarray}
\label{eq:unstr_conv_sol}
\hat w(z) = A e^{z/\delta} + B  e^{-z/\delta} - \frac{2iS}{k \Lambda}  \mbox{   ,   }  \nonumber \\
\hat v(z) = - \frac{1}{ik\delta } \left[ A e^{z/\delta} - B e^{-z/\delta} \right]  \mbox{   ,   }  \nonumber \\
\hat u(z) = \hat u_{\rm cz}(z) - \frac{2}{ik\Lambda\delta } \left[ A e^{z/\delta} - B e^{-z/\delta} \right]  \mbox{   ,   }  \nonumber \\
\hat p(z) = -\frac{2}{ik}  \hat u_{\rm cz}(z)  -\delta \Lambda \left[ A e^{z/\delta} - B e^{-z/\delta} \right]  \mbox{   .   } 
\end{eqnarray}
The constants $A$ and $B$ are integration constants which must be determined
by applying boundary conditions (at $z=1$) and matching conditions (at $z=h$). Note from the $\hat u$-equation that the actual rotation profile 
approaches the imposed (observed) profile 
$\hat u_{\rm cz}$ provided $A$ and $B$ tend
to 0, or when $\Lambda \gg 2$ (in which case $\delta \rightarrow 1/k$). 

\subsubsection{Solution in the radiative zone with stress-free lower boundary, and matching}
\label{sec:unstr_radbc1_ss}

In the radiative region, the equations reduce to 
\begin{eqnarray}
\label{eq:unstr_radbc1_sys}
&&-2 \hat v = \Enu \left(\frac{\dd^2 \hat u}{\dd z^2} - k^2 \hat u \right) \mbox{   ,   }  \nonumber \\
&&2 \hat u = - ik \hat p  \mbox{   ,   }  \nonumber \\
&& \frac{\dd \hat p}{\dd z} = 0 \mbox{   ,   }  \nonumber \\
&& ik \hat v  + \frac{\dd \hat w}{\dd z} = 0 \mbox{   ,   } 
\end{eqnarray}
if we neglect viscous stresses in both $y$ and $z$ components of the 
momentum equation. Note that viscous stresses in the $x-$equation cannot be 
dropped since they are the only force balancing the Coriolis force. 
These equations are easily solved: 
\begin{eqnarray}
\label{eq:unstr_radbc1_sol}
\hat p(z) = p_{\rm rz}  \mbox{   ,   } \nonumber \\
\hat u(z) = - \frac{ik}{2} p_{\rm rz} \mbox{   ,   }  \nonumber \\
\hat v(z) =  - \frac{i E_\nu k^3}{4} p_{\rm rz}  \mbox{   ,   }  \nonumber \\
\hat w(z) =  w_{\rm rz} - \frac{ E_\nu k^4}{4} p_{\rm rz} z \mbox{   ,   } 
\end{eqnarray}
where $p_{\rm rz}$ and $w_{\rm rz}$ are two additional integration constants. 
Here, we recover the standard Taylor-Proudman constraint where in 
the absence diffusion 
or any other stresses, the velocity must be constant along the rotation axis 
(here, $\ez$); in the limit $\Enu \rightarrow 0$, $\hat u(z)$ and $\hat w(z)$ become independent of $z$, while $\hat v(z)\rightarrow 0$.

We are now able to match the solution in the radiative zone to that of the 
convection zone. The two constants $p_{\rm rz}$ and $w_{\rm rz}$ form, together 
with $A$ and $B$, a set of 4 unknown constants which are determined by 
application of boundary and matching conditions. 
Since we have neglected viscous effects in the convection zone, we cannot 
require any boundary or matching condition on the horizontal fluid motions.
On the other hand, we are allowed to impose impermeability $\hat w=0$ 
at the surface ($z=1$) and at the bottom ($z=0$). Moreover, we request the 
continuity of the radial (vertical) velocity and of the pressure 
at the interface ($z=h$). Applying these conditions yields the set of equations
\begin{eqnarray}
\label{eq:unstr_matchbc1_sys}
w_{\rm rz} = 0  \mbox{   ,   }  \nonumber  \\
A e^{h/\delta} + B  e^{-h/\delta} - \frac{2iS}{k \Lambda} = w_{\rm rz} - \frac{\Enu k^4}{4} p_{\rm rz} h  \mbox{   ,   }  \nonumber \\ 
-\frac{2}{ik} U_0(h)  -\delta \Lambda \left[ A e^{h/\delta} - B e^{-h/\delta} \right] = p_{\rm rz} \mbox{   ,   }  \nonumber \\
A e^{1/\delta} + B  e^{-1/\delta} - \frac{2iS}{k \Lambda} = 0  \mbox{   ,   } 
\end{eqnarray}
which have the following solution for $A$ and $B$:
\begin{eqnarray}
\label{eq:unstr_ABbc1_sol}
A =   \frac{ \frac{ \Enu h k^3}{2i} U_0(h) +  \frac{2iS}{k \Lambda}\left[1  -  e ^{(1-h)/\delta}  \left[ 1 + \frac{ E_\nu k^4}{4}h \delta \Lambda \right] \right] }{  e^{h/\delta} \left[ 1 - \frac{ E_\nu k^4}{4}h \delta \Lambda \right]  - e^{(2-h)/\delta} \left[ 1 + \frac{ E_\nu k^4}{4}h \delta \Lambda \right]  } \mbox{   ,   }  \nonumber \\
B = \frac{2iS}{k \Lambda} e ^{1/\delta} - A e^{2/\delta}  \mbox{   .   } 
\end{eqnarray} 
These can be substituted back into (\ref{eq:unstr_conv_sol}) 
to obtain the meridional flow velocities in the convection zone. 
While the exact form of $A$ and $B$ are not particularly informative, 
we note that in the limit $S = 0$ (i.e. the forcing velocity has no 
azimuthal vorticity), both $A$ and $B$ scale as $\Enu$. 
This implies that the amplitude of meridional flows everywhere in the solar
interior scales like $\Enu$ (even in the convection zone). 
The physical interpretation of this somewhat surprising limit is 
discussed in Section \ref{sec:phys_unstrat}, but turns out to be 
of academic interest only (Section \ref{sec:twist}). 

When $S \neq 0$ then $A$ and $B$ are of order $S/k\Lambda$ 
in the convection zone regardless of the Ekman number, 
and respectively tend to
\begin{eqnarray}
A =  \frac{2iS}{k \Lambda} \frac{1  -  e ^{(1-h)/\delta} }{  e^{h/\delta}  - e^{(2-h)/\delta}  } + O(\Enu)  \mbox{   ,   }  \nonumber \\
B = \frac{ 2iS}{  k \Lambda } \frac{1- e^{(h-1)/\delta}  }{ e^{-h/\delta}  -  e^{(h-2)/\delta}  } + O(\Enu) \mbox{   ,   } 
\end{eqnarray} 
as $\Enu \rightarrow 0$. This implies that $\hat w$ is of order $S/k\Lambda$ 
in the convection zone. Since significant flows are locally generated, 
one may reasonably expect a fraction of the forced mass flux
to penetrate into the lower region, especially in this unstratified case. 

Using (\ref{eq:unstr_ABbc1_sol}) in (\ref{eq:unstr_matchbc1_sys}), 
solving for $p_{\rm rz}$, then plugging $p_{\rm rz}$ 
into (\ref{eq:unstr_radbc1_sol}), we find that the general 
expression for $\hat w(z)$ {\it in the radiative zone} $z\in[0,h]$ is 
\begin{equation}
\hat w(z) = - \frac{i E_\nu k^3}{2} \left( U_0(h) +   \frac{ \cosh((1-h)/\delta) - 1 }{\sinh((1-h)/\delta) }    \delta S   \right) z \mbox{   .   } 
\label{eq:unstrat_wrz}
\end{equation}
This implies that only a {\it tiny} fraction of the large 
mass flux circulating in the convection zone 
actually enters the radiative zone. Instead, the system adjusts 
itself in such a way as to ensure that most of 
the meridional flows return above the base of the convection zone. 

We now compare this a priori counter-intuitive\footnote{but a posteriori 
obvious, see Section \ref{sec:phys_unstrat}} analytical result with 
exact numerical solutions of the governing equations.
The numerical solutions were obtained by solving (\ref{eq:maineqs})
for Ro$_{\rm rz} = 0$ (unstratified case), and are uniformly 
calculated in the whole domain (i.e. there is nothing special about the 
interfacial point $z=h$). The boundary conditions used 
are impermeable boundary conditions at the 
top and bottom of the domain for $\hat w$, and stress-free boundary 
conditions for $\hat u$ and $\hat v$. 

\begin{figure}[h]
\centerline{\epsfig{file=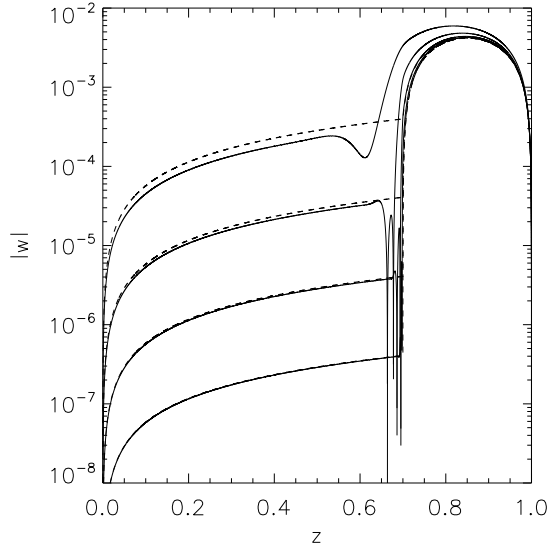,width=8cm}}
\caption{Numerical (solid) and analytical (dashed) solutions for 
$|\hat w(z)|$, in the case of a stress-free bottom boundary. From the uppermost to lowermost curves, 
$E_\nu = 10^{-3}$, $10^{-4}$, $10^{-5}$ and $10^{-6}$ respectively, confirming the analytical
scaling that $\hat w(z) \propto \Enu z$ in the radiative zone while $\hat w(z)$ 
becomes independent of $\Enu$ in the convection zone. These solutions were 
obtained with forcing defined by the parameters $\Delta = 10^{-4}$, $\Lambda=10$, 
$U_0(z) = S(z-h)$ and $S=1$. }
\label{fig:unstr_bc1_wplot1} 
\end{figure}

In Figure \ref{fig:unstr_bc1_wplot1} we compare numerical and 
analytical solutions for $\hat w(z)$, in a case where the forcing 
function parameters are $\Delta = 10^{-4}$, $\Lambda=10$, $U_0(h) = 0$ 
and $S=1$, for four values of the Ekman number. The 
analytical solution is described by equations (\ref{eq:unstr_conv_sol}) 
and (\ref{eq:unstr_ABbc1_sol}) (for the convection zone) 
and (\ref{eq:unstr_radbc1_sol}) (for the radiative zone). As 
$\Enu\rightarrow 0$ the numerical solution approaches the 
analytically derived one, confirming in particular 
that $\hat w(z) \propto z \Enu$ in the radiative zone. The convection zone 
solution is also well-approximated in this case by the analytical formula.

\begin{figure}
\centerline{\epsfig{file=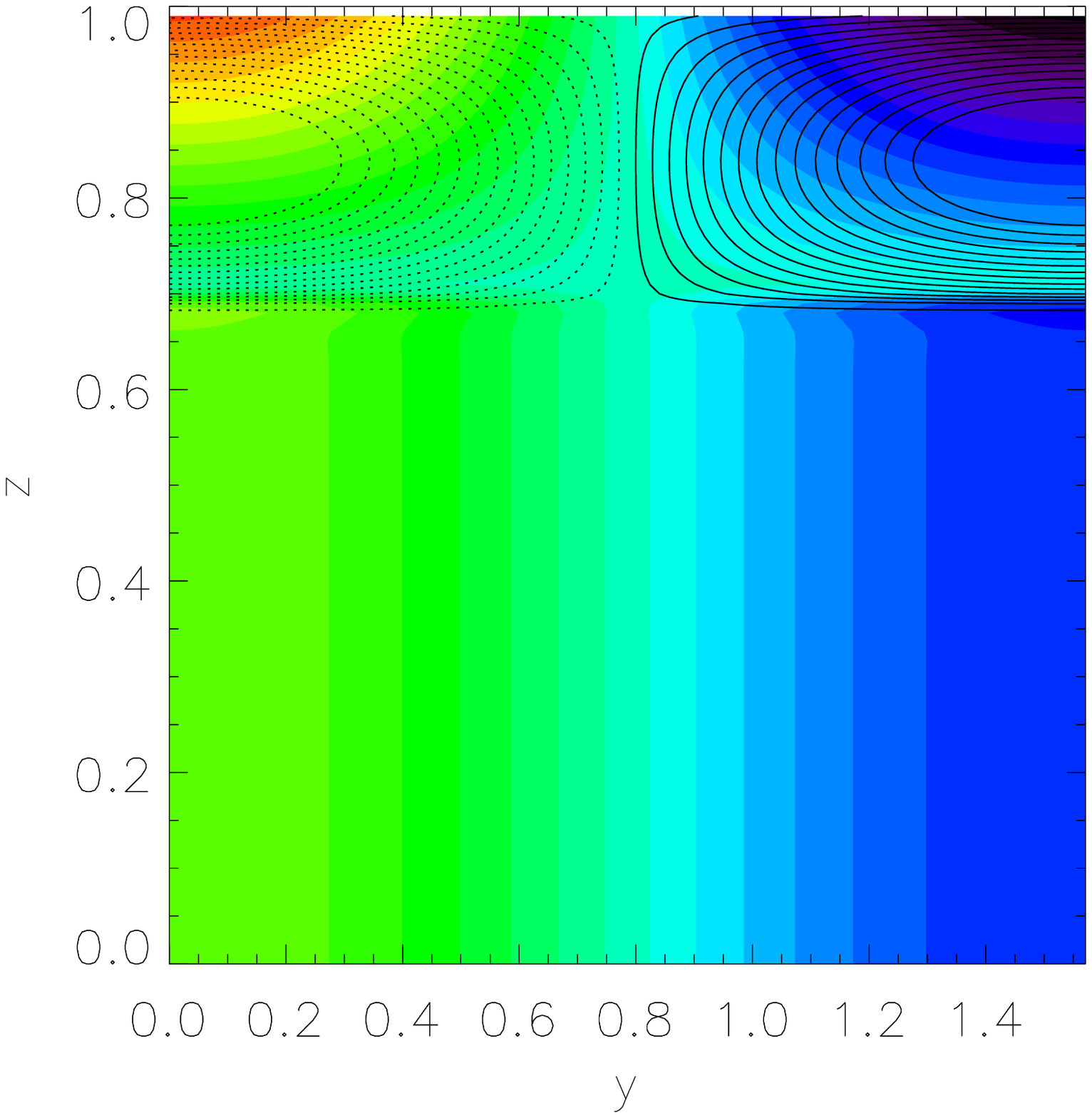,width=8cm}}
\caption{2D visualization of the flow for $E_\nu = 10^{-4}$, in the
  case of a stress-free bottom boundary.  Shown as solid and dotted
  line respectively are linearly spaced streamlines of
  counter-clockwise and clockwise meridional flows.  As predicted, the
  flows appear to return entirely within the convection zone and carry
  a negligible  mass flux into the radiative zone. Meanwhile the
  azimuthal velocity ($\hat u$) as displayed in the filled contours 
  is constant along the rotation axis
  ($z-$axis) below the interface ($z=h=0.7$),  but is strongly sheared
  at the interface. This solution was  obtained with forcing defined
  by the parameters $\Delta = 10^{-4}$, $\Lambda=10$,  $U_0(z) =
  S(z-h)$ and $S=1$, as in Figure \ref{fig:unstr_bc1_wplot1}.  }
\label{fig:unstr_bc1_2Dfield1} 
\end{figure}

A full 2D visualization of the flow for $\Enu = 10^{-4}$ but otherwise 
the same governing parameters is shown in Figure 
~\ref{fig:unstr_bc1_2Dfield1}. This figure illustrates more
clearly the fact that the meridional flows are negligible 
below the interface, and mostly return within the convection zone. Note that given our 
choice of the forcing function $u_{\rm cz}(y,z) \propto \cos(2y)$, the induced Coriolis force does 
not vanish at $y=0$ or $y=\pi$ (the ``poles''). This explains why the meridional flows apparently 
cross the polar axis in this simple model. This is merely a geometric effect:
in a true spherical geometry the forcing azimuthal velocity $u_{\rm cz}(r,\theta)$ would 
be null at the poles, and meridional flows cannot cross the polar axis. More realistic calculations 
in spherical geometry are discussed in Section \ref{sec:spherical}.

\begin{figure}[h]
\centerline{\epsfig{file=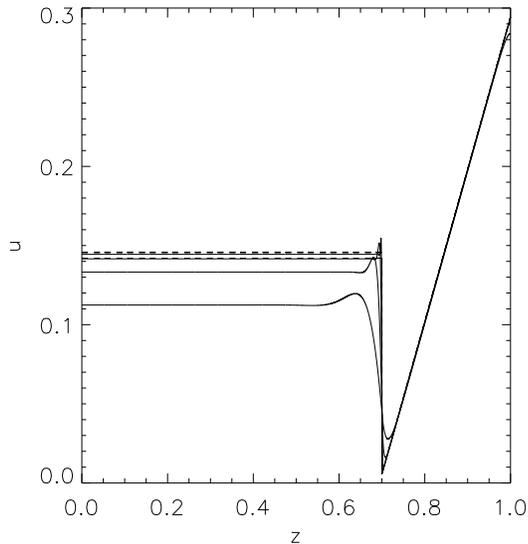,width=8cm}}
\caption{Numerical (solid) and analytical (dashed) solutions for $\hat u(z)$, in the case of a stress-free 
bottom boundary. From the lowermost to uppermost curves, 
$E_\nu = 10^{-3}$, $10^{-4}$, $10^{-5}$ and $10^{-6}$ respectively, confirming 
that $\hat u(z)$ tends to a constant in the radiative zone, while sustaining a finite 
discontinuity at the radiative-convective interface ($z=h=0.7$). 
These solutions were 
obtained with forcing defined by the parameters $\Delta = 10^{-4}$, $\Lambda=10$, 
$U_0(z) = S(z-h)$ and $S=1$, as in Figure \ref{fig:unstr_bc1_wplot1}. }
\label{fig:unstr_bc1_uplot1}
\end{figure}

Finally, it is interesting to note that the analytical solution for
the azimuthal velocity $\hat u(z)$ exhibits a ``discontinuity''
across the base of the convection zone, which tends to
\begin{equation}
\hat u (h^+)-  \hat u (h^-) = \left( \frac{2}{ik\delta \Lambda} 
+ \frac{ik\delta \Lambda}{2} \right) \left( A e^{h/\delta} - B e^{-h/\delta} \right)
\end{equation}
as $\Enu \rightarrow 0$. The numerical solutions of course are continuous,
but the continuity is only assured by the viscosity in the system (in the $y-$ 
direction) and the fact that the overshoot layer depth is finite.
This is shown in Figure \ref{fig:unstr_bc1_uplot1}, together 
with a comparison of the numerical solutions with the analytical solution,
again confirming the analytical approximation derived. 

This highlights another and equally a priori counter-intuitive 
property of the system: the 
value of $u_{\rm rz}$ in the radiative zone is markedly different 
from the imposed
$\hat u_{\rm cz}(h) = U_0(h)$ at the interface:
\begin{equation}
u_{\rm rz} = U_0(h) + \frac{ik\delta\Lambda}{2} \left( Ae^{h/\delta} - B e^{-h/\delta} \right) \mbox{   .} 
\end{equation}
 Hence, even if the imposed differential rotation is 
exactly 0 at the radiative--convective interface (as it is the case 
in the simulation presented in Figure \ref{fig:unstr_bc1_uplot1} since 
$U_0(z) = S(z-h)$), a large-scale latitudinal shear measured by $u_{\rm rz}$ 
may be present in the radiative zone, as illustrated in
Figure \ref{fig:unstr_bc1_2Dfield1}. This shows that 
the propagation of the azimuthal shear into the radiative zone is non-local
(i.e. does not rely on the presence of shear at the interface), 
and is instead communicated by the long-range pressure gradient. 

\subsubsection{Solution in the case of no-slip bottom boundary}
\label{sec:radstrat_noslip}

The stress-free bottom boundary conditions studied in the previous Section
are at first glance the closest to what one may expect in the real Sun, where the 
``bottom'' boundary merely represents the origin of the spherical coordinate system. 
However, let us now explore for completeness (and for further
reasons that will be clarified in the next Section) the case of no-slip bottom boundary 
conditions. 

When the lower boundary is a no-slip boundary, the nature of the solution in 
the whole domain changes. This change is induced by the presence of an Ekman 
boundary layer, which forms near $z=0$. Just above the boundary layer, 
in the bulk of the radiative zone, 
the solution described in \ref{sec:unstr_radbc1_ss} remains valid. 
However, matching the bulk 
solution with the boundary conditions can no longer be done directly; 
one must first solve for the boundary layer dynamics to match the bulk 
solution with the boundary conditions {\it across} the boundary layer. 
This is a standard procedure 
(summarized in Appendix A for completeness), and leads to the well-known 
``Ekman jump'' relationship between the jump in $\hat w(z)$ and the jump 
in $\hat u(z)$ across the boundary layer:
\begin{equation} 
\hat u_{\rm bulk} - \hat  u(0)  = \frac{ 2i}{k} E_\nu^{-1/2} (\hat  w_{\rm bulk} - \hat  w(0)) \mbox{   .  }
\end{equation}

By impermeability, $\hat w(0) = 0$. Moreover, by assuming that the
total angular momentum of the lower boundary is the same as that of
the convection zone, we require that $\hat u(0) = 0$. Meanwhile, $\hat
u_{\rm bulk} = u_{\rm rz}$ and $\hat w_{\rm bulk} = w_{\rm rz}$ in the
notation of equation (\ref{eq:unstr_radbc1_sol}). So finally, for
no-slip boundary conditions, we simply replace the impermeability
condition ($ w_{\rm rz} = 0$) in (\ref{eq:unstr_matchbc1_sys}) by 
\begin{equation} 
u_{\rm rz} = \frac{ 2i}{k} E_\nu^{-1/2} w_{\rm rz}   \mbox{   ,  }
\end{equation}
and solve for the unknown constants $A$, $B$, $ w_{\rm rz} $ 
and $p_{\rm rz}$ as before.

The exact expressions for the resulting integration constants $A$ and $B$ are 
now slightly different from those given in (\ref{eq:unstr_ABbc1_sol}), but
are without particular interest. However, it can be shown that 
they have the same limit as 
in the stress-free case as $\Enu \rightarrow 0$ (with $S \neq 0$).
This implies that the meridional flows driven {\it within the convection zone}, 
in the limit $\Enu \rightarrow 0$, and with $S \neq 0$, 
are independent of the boundary condition 
selected at the bottom of the radiative zone. However, we now have
the following expression for $\hat w(z)$ in the bulk of the radiative zone: 
\begin{eqnarray}
\hat w(z)  &=& -\left( \frac{k^2}{4} E_\nu^{1/2} + \frac{E_\nu k^4}{4} z \right) 
p_{\rm rz} \mbox{   ,  }\nonumber \\
&=&  -\frac{ik}{2} E_\nu^{1/2} \left( U_0(h) +   \frac{ \cosh((1-h)/\delta) - 1 }{\sinh((1-h)/\delta) }    \delta S   \right) + O(\Enu) z \mbox{   ,  }
\end{eqnarray}
which has one fundamental consequence: the amplitude of the flows 
allowed to penetrate into the radiative zone is now of order $E_\nu^{1/2}$ 
instead of being $O(\Enu)$. This particular statement is actually true even 
if $S=0$, although in that case both convection zone and radiative 
zone flows scale with $E_\nu^{1/2}$. 

\begin{figure}[h]
\centerline{\epsfig{file=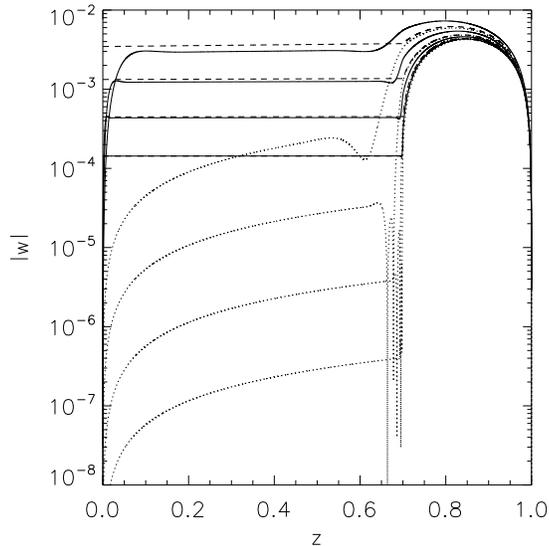,width=8cm}}
\caption{Numerical (solid) and analytical (dashed) solutions for 
$|\hat w(z)|$, in the case of a no-slip bottom boundary. From the uppermost to lowermost curves 
(as seen in the radiative zone), 
$E_\nu = 10^{-3}$, $10^{-4}$, $10^{-5}$ and $10^{-6}$ respectively, confirming the analytical
scaling that $\hat w(z) \propto \Enu^{1/2}$ in the radiative zone while $\hat w(z)$ 
becomes independent of $\Enu$ in the convection zone. These solutions were 
obtained with forcing defined by the parameters $\Delta = 10^{-4}$, $\Lambda=10$, 
$U_0(z) = S(z-h)$ and $S=1$, as in Figure \ref{fig:unstr_bc1_wplot1}. 
For comparison, the 
previous simulations with stress-free bottom boundary, for the same parameters, are shown as dotted lines.}
\label{fig:unstr_bc2_wplot1} 
\end{figure}

\begin{figure}
\centerline{\epsfig{file=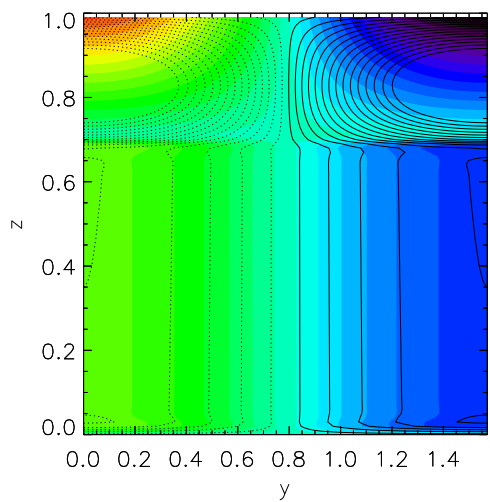,width=8cm}}
\caption{The same as for Figure \ref{fig:unstr_bc1_2Dfield1} but for no-slip boundary conditions. The Ekman layer near the lower boundary is clearly visible. For ease of comparison, the same streamlines are shown in the two plots. The two figures illustrates how the nature of the lower boundary condition influences the mass flux through the radiative zone. 
}
\label{fig:unstr_bc2_2Dfield1} 
\end{figure}

Figure \ref{fig:unstr_bc2_wplot1} 
shows a comparison between the approximate analytical formula
and the numerical solution for the same simulations as in Figure 
\ref{fig:unstr_bc1_wplot1}, but now using no-slip bottom boundary conditions. 
For ease of comparison, the results from the stress-free numerical simulations
(for exactly the same parameters) have also been drawn, highlighting the much 
larger amplitude of the meridional flows down-welling into the radiative zone in the no-slip case, and their scaling with $E_\nu^{1/2}$.
Figure \ref{fig:unstr_bc2_2Dfield1} shows an equivalent 2D rendition of the solution,
and illustrates the presence of large-scale mixing in the bulk of the 
radiative zone when the bottom-boundary is no-slip. 

\subsubsection{Physical interpretation}
\label{sec:phys_unstrat}

The various sets of solutions derived above can be physically understood 
in the following way.
Let us first discuss the solution in the convection zone. 
In the limit where $u_{\rm cz}(y,z)$ is independent of $z$ 
(equivalently, $S=0$), the azimuthal ($x-$) component of the vorticity of the 
forcing is zero. In that case there is  
no injection of $x-$vorticity into the 
system aside from that induced in the viscous boundary layers, and the 
amplitude of the meridional flows generated in the convection zone scales
with $\Enu$. This limit is somewhat academic in the case of the Sun, 
however given the observed rotation profile (see also Section \ref{sec:twist}). 
When $S\neq0$, the amplitude of 
the induced meridional flows in the convection zone 
scales linearly with $S$ and is independent of viscosity. 

In the radiative zone, the Taylor-Proudman constraint enforces invariance 
of the flow velocities along the rotation axis, 
except in regions where other forces balance the Coriolis force. 
In the non-magnetic, unstratified situation discussed in the two previous 
sections, the only agent capable of 
breaking the Taylor-Proudman constraint are viscous stresses, which are only 
significant in two thin boundary layers: 
one right below the convection zone and the other one 
near the bottom boundary. {\it These two layers are the only regions 
where flows down-welling into the radiative zone are allowed to return 
to the convection zone}. The question then remains of what fraction of 
the mass flux entering the radiative zone returns within the upper 
Ekman layer, and what fraction returns within the lower Ekman layer. The 
latter, of course, permits large-scale mixing within the radiative zone. 

In the first case studied, the bottom boundary was 
chosen to be stress-free. This naturally suppresses the lower 
viscous boundary layer so that the only place where flows are allowed to 
return  is at the radiative-convective 
interface. As a result, only a tiny fraction of the mass flux penetrates 
below $z=h$, and the turnover time of the remaining flows within the 
radiative zone
is limited to a viscous timescale of the order of $1 / \Enu \Omega_\odot $. 

Following this reasoning, we expect and indeed find 
quite a different behavior when the bottom boundary is no-slip. In that case
viscous stresses within the lower boundary layer break the Taylor-Proudman constraint and 
allow a non-zero mass flux (of order $\Enu^{1/2}$) to return near $z=0$. 
This flow then mixes the entire radiative zone as well, with an 
overall turnover time of order of $1 / \Enu^{1/2} \Omega_\odot $ 
(in dimensional units). 

To summarize, in this unstratified 
steady-state situation, the amount of mixing induced within the 
radiative zone by convective zone flows depends (of course) on the 
amplitude of the convection zone forcing, but also on the existence of
a mechanism to break the Taylor-Proudman constraint somewhere 
within the radiative zone. That mechanism is needed in order to allow
down-welling flows to return to the convection zone. But more 
crucially, this phenomenon implies that the dynamics of the lower boundary 
layer entirely control the mass flux through the system.

Here, we studied the case of viscous stresses only. One can rightfully 
argue that there are no expected ``solid'' boundaries in a stellar 
interior and that the overall behavior of the system should be 
closer to the one discussed in the stress-free case than the no-slip case.
{\it However, we chose here to study viscous stresses simply because they 
are the easiest available example.} In real stars viscous stresses 
are likely to be negligible compared with a variety 
of other possible stresses: turbulent stresses at the interface with 
another convection zone, magnetic forces, etc. Nevertheless, 
these stresses will play a similar role in allowing flows to mix the 
radiative zone if they become comparable in amplitude with the Coriolis force, 
and help break the Taylor-Proudman constraint. This issue is discussed in more 
detail in Section \ref{sec:disc}.
 
\subsubsection{The thickness of the overshoot layer}
\label{sec:twist}

Before moving on to the more realistic stratified case, note that
this unstratified system holds one final subtlety. In all 
simulations presented 
earlier, the overshoot layer depth was selected to be very small -- and in 
particular, smaller than the Ekman layer thickness. In that case,
the transition in the forcing at the base of the 
convection zone is indeed close to being a discontinuity, and the 
analytical solutions presented in Sections \ref{sec:unstr_conv_ss} and \ref{sec:unstr_radbc1_ss} are a good fit to the true numerical solution.

In the Sun, the overshoot layer depth 
$\Delta$ is arguably always thicker than an Ekman lengthscale $\Enu^{1/2}$.
When this happens, the solution ``knows'' about  
the exact shape of the forcing function within the transition region,
and therefore depends on it.
This limit turns out to be rather difficult to study analytically, and since
in the case of the Sun we do not know the actual profile of $\tau^{-1}(z)$,
there is little point in exercise anyway. 

\begin{figure}[h]
\centerline{\epsfig{file=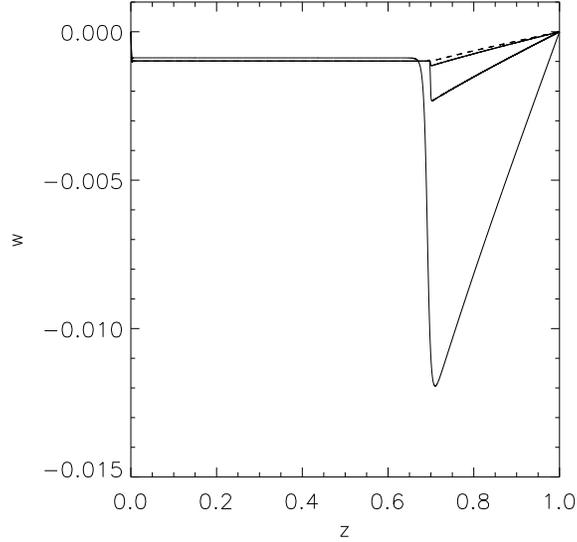,width=8cm}}
\caption{Comparison of numerical simulations (solid lines) with analytical prediction (dashed line) for a no-slip bottom boundary, for $\Enu = 10^{-6}$, and forcing functions defined with $\Lambda=10$, $S=0$ and $U_0(h) = 1$. The three numerical solutions are obtained for various values of the overshoot layer depth: 
from lowermost to uppermost curves (as seen in the convective zone), $\Delta = 10 \Enu^{1/2}$, $\Enu^{1/2} $, and $0.1\Enu^{1/2}$. The analytical solution assumes an infinitely thin overshoot layer and is therefore independent of $\Delta$. Note that the analytical solution in the convection zone is only a good approximation to the true solution if $\Delta \ll \Enu^{1/2}$. The overall scalings in the radiative zone, however, are preserved.
}
\label{fig:varydelta} 
\end{figure}

We can explore the behavior of the system numerically, however, for the 
profile $\tau^{-1}(z)$ discussed in equation (\ref{eq:tau_profile}), in the limit where $\Delta > \Enu^{1/2}$. The example for which this effect matters the 
most is the somewhat academic limit where $S=0$ in the convection zone, but 
$U_0(h) \neq 0$. In this case, the asymptotic analysis predicts that 
the meridional flow amplitudes are $O(\Enu^{1/2})$ in both the convection zone 
and in the radiative zone for the no-slip case. 
We see in Figure \ref{fig:varydelta} that this 
is indeed the case in simulations where
$\Delta \ll \Enu^{1/2}$. However, when the overshoot thickness is 
progressively increased and becomes larger than the Ekman layer thickness, 
the amplitude of the meridional circulation 
in the convection zone is no longer $O(\Enu^{1/2})$ but much larger. 
Meanwhile, the scaling of the radiative zone solutions with $\Enu$ 
remain qualitatively correct. The difference with the 
analytical solution in the convection zone can simply be attributed 
to the fact that when the system knows about the shear {\it within} 
the overshoot layer the limit $S=0$ is no longer relevant. 

\subsection{The stratified case}
\label{sec:strat}

While the previous section provides interesting insight into the 
problem, notably on the role of the Taylor-Proudman constraint, 
we now move to the more realistic situation where
stratification plays a role in the flow dynamics. 
In this section, we generalize our Cartesian study to take into 
account the stratification of the lower region (Ro$_{\rm rz} \neq 0$).
For this purpose, we go back to studying the full system of equations 
(\ref{eq:maineqs}). As before, we first find approximate analytical 
solutions to derive the overall scaling of the solutions with 
governing parameters, and then compare them to the full numerical 
solutions of (\ref{eq:maineqs}). The analytical solutions are obtained
by solving the system in the convective region and radiative region 
separately, and matching them at $z=h$.

\subsubsection{Convection zone solution}

The equations in the convection zone are now given by 
\begin{eqnarray}
&&-2 \hat v = - \Lambda(\hat u-\hat u_{\rm cz}) \mbox{   ,   }  \nonumber \\
&&2 \hat u = - ik \hat p  - \Lambda \hat v  \mbox{   ,   }  \nonumber \\
&&0 = - \frac{\dd \hat p}{\dd z} - \Lambda \hat w  \mbox{   ,   }  \nonumber \\
&& \hat w = D_0 \left( \frac{\dd^2 \hat T}{\dd z^2} - k^2 \hat T \right) \mbox{   ,   }  \nonumber \\
&& ik \hat v  + \frac{\dd \hat w}{\dd z} = 0 \mbox{   ,   } 
\end{eqnarray}
where we have assumed that $D_0 \gg \Enu/{\rm Pr}$. 
Eliminating variables one by one yields the same equation for 
$\hat w(z)$ in the convection zone as before 
(\ref{eq:unstr_conv_weq}), as well as 
a Poisson equation for $\hat T$ once $\hat w$ is known. 
The solutions are then as (\ref{eq:unstr_conv_sol}), together with
\begin{equation}
\hat T(z) = T_0 e^{kz} + T_1 e^{-kz} + \frac{\delta^2 \left[ A e^{z/\delta} +  B e^{-z/\delta} \right]}{D_0 ( 1- \delta^2 k^2)}  + \frac{2iS}{k^3 \Lambda D_0}\mbox{   ,  }
\label{eq:tconv}
\end{equation}
where the integration constants $T_0$ and $T_1$ remain to be determined. 
For the sake of analytical simplicity, we will assume that $D_0 \gg 1$ 
in all that follows (i.e. very large thermal diffusivity in the convection 
zone), and thus neglect the third and fourth terms in (\ref{eq:tconv}). 
This limit is relevant for the Sun. 

\subsubsection{Radiative zone solution}
\label{sec:stratrad}

The radiative zone equations are now
\begin{eqnarray}
 - 2\hat v =  E_\nu \left( \frac{\dd^2 \hat u}{\dd z^2} - k^2 \hat u \right)   \mbox{   ,   } \nonumber \\
 2 \hat u = - ik \hat p \mbox{   ,   }  \nonumber \\
 0 = - \frac{ \dd \hat p}{\dd z}  + {\rm Ro}^2_{\rm rz}(z)\hat T \mbox{   ,   } \nonumber \\
\hat w = \frac{\Enu}{{\rm Pr}} \left( \frac{\dd^2 \hat T}{\dd z^2} - k^2 \hat T \right) \mbox{   ,   } \nonumber \\
ik \hat v + \frac{\dd \hat w}{\dd z} =0 \mbox{   ,   } 
\end{eqnarray}
and can be combined to yield 
\begin{equation}
\frac{\dd^4 \hat u}{\dd z^4} - k^2 \left(1 + \frac{{\rm Pr Ro_{\rm rz}}^2 }{4}  \right) \frac{\dd^2 \hat u}{\dd z^2} + k^4 \frac{{\rm Pr Ro_{\rm rz}}^2 }{4} \hat u = 0\mbox{   ,   } 
\end{equation}
and similarly for $\hat T$. The characteristic polynomial is 
\begin{equation}
(\lambda^2 - k^2 ) \left( \lambda^2  - \frac{{\rm Pr Ro_{\rm rz}}^2 }{4} k^2 \right) = 0 \mbox{   ,   } 
\end{equation}
with solutions 
\begin{eqnarray}
\pm \lambda_1 = \pm k \mbox{   ,   } \nonumber \\
\pm \lambda_2 = \pm \sqrt{\rm Pr} \frac{\rm Ro_{\rm rz}}{2} k\mbox{   .   } 
\label{eq:lambda2}
\end{eqnarray} 
These solutions are the same as those presented in Paper I, 
and will be referred to as the ``global-scale'' mode and the thermo-viscous 
mode respectively. Note that here, $\lambda_2$ corresponds to $k_2$ in Paper I.

In this steady-state study, the quantity $\lambda_2$ summarizes 
the effect of stratification. It is important to note that it contains
information about the rotation rate of the star as well as the 
Prandtl number, in addition to the buoyancy frequency. 
If $\lambda_2 \ll 1$, then the thermo-viscous mode essentially spans 
the whole domain: the system appears to be ``unstratified'', and 
is again dominated by the Taylor-Proudman constraint. On the 
other hand, if $\lambda_2 \gg 1$ then the flows only penetrate 
into the radiative zone within a small thermo-viscous boundary 
layer of thickness $1/\lambda_2$ as a result of the strong 
stratification of the system. The Taylor-Proudman constraint is irrelevant in 
this limit, since the magnitude of the buoyancy force is much larger
than that of the Coriolis force. 

The calculation above was made in the limit 
where the viscous terms in the latitudinal 
and radial components of the momentum equation are discarded. Paper I shows
that two additional Ekman modes are also present if they are instead kept. 
By analogy with the unstratified case, we expect that these Ekman modes
do not influence the solution for stress-free 
boundary conditions, but that additional 
care must be taken for no-slip boundary conditions. 

Note that the equation for $\hat w$ instead simplifies to 
\begin{equation}
\frac{\dd^2 \hat w}{\dd z^2} = k^2 \frac{{\rm Pr Ro_{\rm rz}}^2 }{4} \hat w \mbox{   ,   } 
\end{equation}
and similarly for $\hat v$ (i.e. both equations are only second order in $z$, and only contain the thermo-viscous mode). The radiative zone ($z\in [0,h]$) solutions are now
\begin{eqnarray}
&&\hat u(z) = u_1 e^{k z} + u_2 e^{-k z} + u_3 e^{\lambda_2 z} + u_4 e^{-\lambda_2 z} \mbox{   ,   } \nonumber \\
&&\hat v(z) = \frac{E_\nu}{2}  (k^2 - \lambda_2^2 ) \left[ u_3 e^{\lambda_2 z} + u_4 e^{-\lambda_2 z} \right]\mbox{   ,   }  \nonumber \\
&&\hat w(z) = - ik \frac{E_\nu}{2}  \frac{(k^2 - \lambda_2^2 )}{\lambda_2} \left[ u_3 e^{\lambda_2 z} - u_4 e^{-\lambda_2 z} \right] \mbox{   ,   } \nonumber \\
&& \hat p(z) = -\frac{2}{ik}\left[ u_1 e^{k z} + u_2 e^{-kz} + u_3 e^{\lambda_2 z} + u_4 e^{-\lambda_2 z} \right]  \mbox{   ,   } \nonumber \\
&& \hat T(z)= -  \frac{2}{ik {\rm Ro}^2_{\rm rz}} \left[ k u_1 e^{k z} -ku_2 e^{-k z} + \lambda_2 u_3 e^{\lambda_2 z} - \lambda_2 u_4 e^{-\lambda_2 z} \right]\mbox{   .   } 
\end{eqnarray}
where the 4 constants $\{u_i\}_{i=1,4}$ are integration constants, to be determined. 

\subsubsection{The stratified stress-free case}
\label{sec:match_strat_stressfree}

We now proceed to match the solutions in the two regions, assuming stress-free boundary conditions near the lower boundary. Since there are in total 8 unknown constants (including $A$, $B$, $T_0$ and $T_1$ from the convection zone solution and $\{u_i\}_{i=1,4}$ from the radiative zone solution), we need a total of 8 matching and boundary conditions.

At the lower boundary ($z=0$) we take $\hat w = \dd \hat u/\dd z = 0$;
this condition in turn implies that $\hat T = 0$. At the surface
($z=1$),  we take as before $\hat w = 0$, and select in addition $\hat
T = 0$.  We then need 4 matching conditions across the interface:
these are given  by the continuity of $\hat w$, $\hat p$, $\hat T$ and
$\dd \hat T/\dd z$.  Note that it is important to resist the
temptation of requiring the  continuity of $\hat v$, since viscous
stresses have been neglected in the analytical treatment of the $y-$
component of the momentum equation in both radiative and convective
zones.  Moreover, we know that in the unstratified limit, $\hat u$
actually  becomes discontinuous at the  interface in the limit $\Enu
\rightarrow 0$. Since we expect the stratified solution to tend to the
unstratified one uniformly as Ro$_{\rm rz} \rightarrow 0$, 
we cannot require the continuity
of $\hat u$ at the interface\footnote{a fact
which is again only obvious in hindsight}. 

The equations and resulting solutions
for the integration constants are  fairly complicated. The most
important ones are reported in  the Appendix B for completeness, and
are used to justify mathematically the following statements:
\begin{itemize}
\item In the limit of Ro$_{\rm rz}$ $\rightarrow 0$, we find as expected
that the solutions uniformly tend to the unstratified solution summarized
in equations (\ref{eq:unstr_conv_sol}), (\ref{eq:unstr_radbc1_sol}), and
(\ref{eq:unstr_ABbc1_sol}). Indeed, in that case $\lambda_2
\rightarrow 0$ and the  thermo-viscous solution spans the whole 
radiative zone (mathematically, it tends to the linear solution 
found in the unstratified case). 
\item In the strongly stratified case (defined as $\lambda_2 \gg k$), as
described earlier, $\hat w$ in the radiative zone decays exponentially with
depth on a lengthscale $1/\lambda_2$, with an amplitude which  scales
as $\Enu / \lambda_2$. The flows are therefore very strongly
suppressed, and return to the convection zone within a small thermo-viscous
layer. Note that  $\Enu / \lambda_2 = {\rm Ra}^{-1/2}$ where Ra is the 
usually defined Rayleigh number. 
\end{itemize}

\begin{figure}[h]
\centerline{\epsfig{file=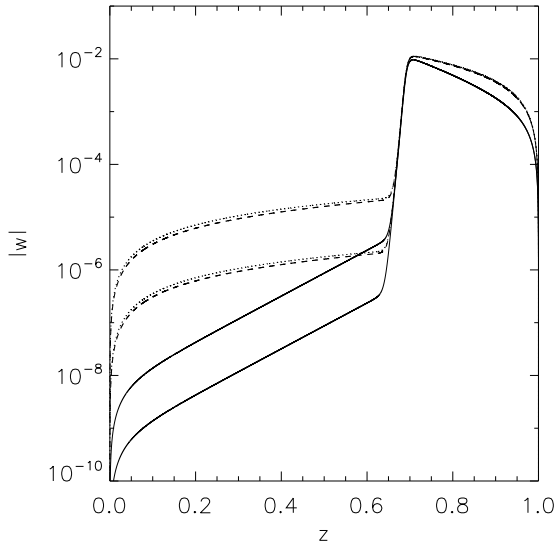,width=8cm}}
\caption{Numerical solutions of (\ref{eq:maineqs}) with the
  following parameters:  $\Delta = 0.01$, $\Lambda = 10$, $S=0$,
  $U=1$, Pr $=0.01$ and $D_0 = 10$. Stress-free bottom boundary conditions 
  are used. The solid lines correspond to the
  ``strongly'' stratified case with  Ro$_{\rm rz}$ = $10$, 
  with $\Enu = 10^{-5}$
  and $10^{-6}$ for the top  and bottom curves respectively. The
  dashed lines correspond to the ``weakly''  stratified case, with  Ro$_{\rm rz}  = 1$, with $\Enu = 10^{-5}$ and $10^{-6}$ for the top  and bottom
  curves respectively. Note that for $k=2$, $\lambda_2$ is simply
  equal to Pr$^{1/2}$Ro$_{\rm rz}$. For comparison, the unstratified
  case (Ro$_{\rm rz}$ = 0) is shown as dotted lines. 
  At these parameters and with 
  these boundary conditions, Ro$_{\rm rz}$ = 1 already belongs to the 
  weakly stratified limit. }
\label{fig:stratunstrat}
\end{figure}

The two limits are illustrated in Figure \ref{fig:stratunstrat}, which
shows the numerical solution to (\ref{eq:maineqs}) for two
values of the Rossby number Ro$_{\rm rz}$, but otherwise identical
parameters. In the strongly stratified limit ($\lambda_2 = 10$, using
Pr = 0.01 and Ro$_{\rm rz} = 10^2$) we see that the solution decays
exponentially below the interface, with an amplitude which scales as
$\Enu/\lambda_2$ as predicted analytically. In the weakly stratified case 
($\lambda_2 = 0.1$, using Pr = 0.01 and Ro$_{\rm rz} = 1$) the 
solution tends to the unstratified limit and scales as $\Enu z$.

\subsubsection{Matching in the no-slip case}

By analogy with the previous section, we expect to recover 
the unstratified limit when $\lambda_2 \rightarrow 0$, so
that $\hat w(z) \propto \Enu^{1/2}$ in this no-slip case.
In the strongly stratified limit
on the other hand, the amplitude of the flows decays 
exponentially with depth below the
interface as a result of the thermo-viscous mode and 
is negligible by the time they reach the lower
boundary. In that case, 
we do not expect the applied lower boundary conditions to 
affect the solution, so that the scalings found in the 
strongly stratified limit with {\it stress-free} boundary conditions 
should still apply: $\hat w(z)\propto \Enu/\lambda_2$.  

\begin{figure}[h]
\centerline{\epsfig{file=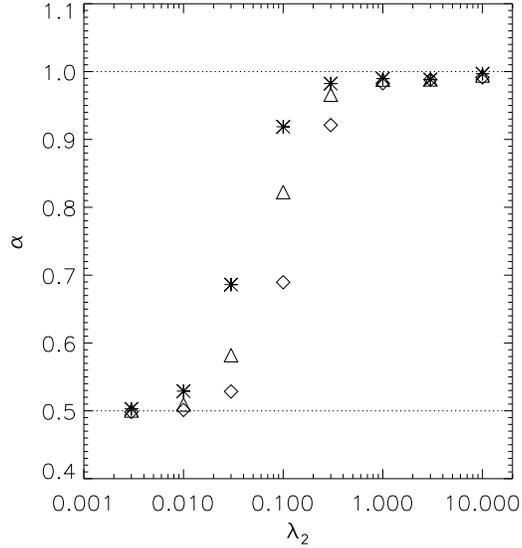,width=8cm}}
\caption{This figure shows the power $\alpha$ in the expression  $\hat
  w \propto \Enu^\alpha$, as a function of $\lambda_2$ (see main text for detail). In the weakly stratified limit, $\alpha
  \rightarrow 1/2$ while in the strongly stratified limit $\alpha
  \rightarrow 1$ as predicted analytically. This calculation was done for no-slip boundary conditions, and the following parameters were held constant: $\Delta = 0.01$, $\Lambda = 10$,
$S=1$, $U_0(h)=1$, Pr $=0.01$ and $D_0 = 10$  }
\label{fig:ratio} 
\end{figure}

These statements are verified in Figure
\ref{fig:ratio}. There, we show the results of a series of numerical 
experiments for no-slip boundary conditions where we extracted 
from the simulations the power $\alpha$ in the expression
$\hat w \propto \Enu^\alpha$, and plotted it 
as a function of stratification ($\lambda_2$). To do this, 
we integrated the solutions to equations (\ref{eq:maineqs}) 
for the following parameters: $\Delta = 0.01$, $\Lambda = 10$,
$S=1$, $U_0(h)=1$, Pr $=0.01$ and $D_0 = 10$ and 
calculated $\hat w(z=0.5)$ for 4 values of $\Enu$:
$10^{-6}$, $10^{-7}$, $10^{-8}$ and $10^{-9}$. We estimated $\alpha$
by calculating the quantity 
\begin{equation}
\alpha = \log_{10}\frac{ \hat
    w(z=0.5,\Enu=10^{-6})}{ \hat w(z=0.5,\Enu=10^{-7})}
\end{equation}
 for the ($\Enu = 10^{-6}$, $\Enu =
  10^{-7}$) pair (diamond
  symbols) and similarly for the pairs ($\Enu = 10^{-7}$, $\Enu =
  10^{-8}$) (triangular symbols) and ($\Enu = 10^{-8}$, $\Enu =
  10^{-9}$) (star symbols).
In the
weakly stratified limit ($\lambda_2 \rightarrow 0$), 
we find that $\alpha \rightarrow 1/2$  while
in the strongly stratified limit ($\lambda_2 \gg 1$), 
$\alpha \rightarrow 1$, thus confirming our analysis. The
transition  between the two regimes appears to occur for slightly
lower-than expected  values of $\lambda_2$, namely 0.1 instead of 1.

A final summary of our findings for the stratified case
together with its implications for mixing between the solar
convection zone and the radiative interior, is deferred to Section 
\ref{sec:disc}. There, we also discuss the consequences in terms of 
mixing in other stars. But first, we complete the study by 
releasing some of the simplifying assumptions made, and moving to more
realistic numerical solutions to confirm our simple Cartesian analysis. 

\section{A ``solar'' model}
\label{sec:spherical}

\begin{figure}[h]
\centerline{\epsfig{file=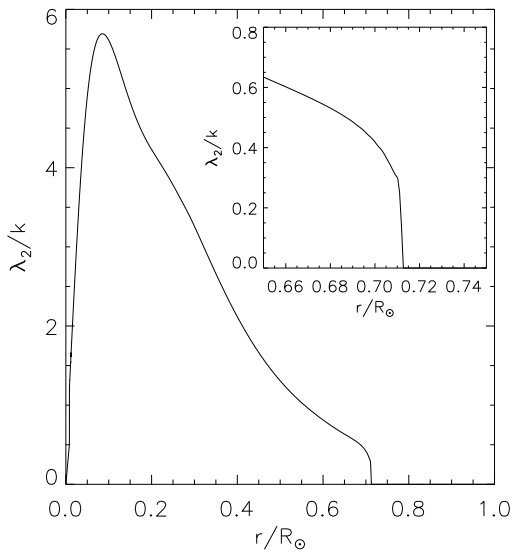,width=8cm}}
\caption{Variation of $\lambda_2/k =$Pr$^{0.5}N/\Omega_\odot$ in the Sun, 
as determined from Model S of Christensen-Dalsgaard {\it et al.} (1996). The 
Prandtl number Pr is calculated using Model S together with the formulae 
provided by Gough (2007) for the microscopic values of the viscosity $\hat \nu$
and the thermal conductivity $\kappa_T$ (see also Garaud \& Garaud, 2008). 
The inset zooms into the region near the base of the convection zone, 
which is the only region of the radiative zone
where $\lambda_2 \le 1$ (aside from $r\rightarrow 0$).} 
 \label{fig:lambda2} 
\end{figure}

In this section we improve on the Cartesian analysis by moving to 
a spherical radiative--convective model. The calculations are thus
performed in an axisymmetric 
spherical shell, with the outer radius $r_{\rm out}$ 
selected to be near the solar surface, and the inner radius $r_{\rm in}$ 
somewhere within the radiative interior. This enables us to 
gain a better understanding of the effects of the geometry of the system 
on the spatial structure of the flows generated. 
In addition, we use more realistic input physics in particular in 
terms of the background stratification, and no longer use the 
Boussinesq approximation for the equation of state. We expect that 
the overall scalings derived in Section \ref{sec:cart} still 
adequately describe 
the flow amplitudes in this new calculation. However, the use of a 
more realistic background stratification adds an additional complication 
to the problem: the background temperature/density gradients are no longer
constant, so that the measure of stratification  $\lambda_2$ 
varies with radius (see Figure \ref{fig:lambda2}, 
for an estimate of $\lambda_2$ in the Sun). 
This aim of this section is therefore to study the impact of both 
geometry and non-uniform stratification on the system dynamics. 

\subsection{Description of the model}

The spherical model used is analogous to the radiative-zone-only 
model presented in Paper I and 
described in detail (including the magnetic case) by Garaud \& Garaud (2008). 
The salient points are repeated here for completeness, together with the 
added modifications made to include the ``convective'' region.

We consider a spherical coordinate system $(r,\theta,\phi)$ where the polar
axis is aligned with rotation axis of the Sun. The background 
state is assumed to be spherically symmetric and in hydrostatic
equilibrium. The background thermodynamical quantities such as 
density, pressure, temperature and entropy are denoted with bars 
(as $\bar{\rho}(r)$, $\bar{p}(r)$, 
$\bar{T}(r)$ and $\bar{s}(r)$ respectively), and extracted from 
the standard solar model of Christensen-Dalsgaard {\it et al.} (1996).
Perturbations to this background induced by the velocity field 
$\bu = (u_r, u_\theta, u_\phi )$ 
are denoted with tildes. In the frame 
of reference rotating with angular velocity $\Omega_\odot$, 
in a steady state, the linearized perturbation equations become
\begin{eqnarray}
&&  \div(\bar{\rho} \bu) = 0 \mbox{   ,   }\nonumber  \\
&& 2\bar{\rho} {\bf \Omega}_\odot \ez \times \bu   =
- \grad \tilde{p} + \tilde{\rho} {\bf g}
+  f \div \Pi - \bar{\rho} \Omega_\odot \frac{\bu - \bu_{\rm cz}}{\tau(r)} \mbox{   ,   } \nonumber  \\
&& \frac{\bar \rho \bar c_{\rm p} \bar T \bar N^2}{g} u_r 
= \div\left[ (f\bar{k}_T + R^2_\odot\Omega_\odot D(r)) \grad \tilde{T}\right]  \mbox{   ,   } \nonumber  \\
&& \frac{\tilde{p}}{\bar{p}} = \frac{\tilde{\rho}}{\bar{\rho}} 
+  \frac{\tilde{T}}{\bar{T}} \mbox{   ,   }
\label{eq:global}
\end{eqnarray}
where $\bar c_{\rm p}$ is the specific heat at constant pressure, 
$\bar{k}_T(r) = \bar{\rho} \bar c_{\rm p} \bar \kappa_T $ 
is the thermal conductivity in the solar interior, 
$\Pi$ is the viscous stress tensor (which depends on the background viscosity 
$\bar \nu$) and $\bg = - g(r) \er$ is gravity. 
Note that this set of equations is given in dimensional form
here, although the numerical algorithm used further casts them 
into a non-dimensional form. Also note that both diffusion terms 
(viscous diffusion and heat diffusion) have been 
multiplied by the same factor $f$. This enables us to vary the 
effective Ekman number 
$\Enu(r) = f\Enu^\odot = f \bar \nu(r)/R^2_\odot \Omega_\odot$ 
while maintaining a solar Prandtl number at every radial position. As a result,
the quantity $\lambda_2$ used in the simulations 
and represented in Figure \ref{fig:lambda2} is 
the true solar value (except where specifically mentioned). 

As in the Cartesian case, we
model the dynamical effect of turbulent convection in the 
convection zone through a relaxation to 
the observed profile in the momentum equation, and a turbulent diffusion 
in the thermal energy equation. The expressions for the non-dimensional
quantities $\tau(r)$ and $D(r)$ 
are the same as in equations (\ref{eq:tau_profile}) and (\ref{eq:D_profile}) 
with $z$ replaced by $r/R_\odot$, and $h = 0.713$ instead of $h=0.7$ 
(Christensen-Dalsgaard {\it et al.} 1996). In what follows, we take 
$\Delta$ to be 0.01 (i.e. the overshoot layer depth is 1\% of the solar 
radius) although the choice of $\Delta$ has 
little influence on the scalings derived. 
The rotation profile in the convection zone $\bu_{\rm cz}$ is 
selected to be
\begin{equation}
\bu_{\rm cz}(r,\theta) = r \sin\theta \Omega_{\rm cz}(\theta) \mbox{  } \ephi \mbox{   , }
\end{equation}
where 
\begin{equation}
\Omega_{\rm cz}(\theta) = \Omega_{\rm eq} \left( 1 - a_2 \cos^2\theta - a_4  \cos^4\theta \right)\mbox{   , }
\end{equation}
with 
\begin{eqnarray}
&& a_2 = 0.17 \mbox{   , } a_4 = 0.08 \mbox{   , } \nonumber \\
&& \frac{\Omega_{\rm eq}}{2\pi} = 463 \mbox{   nHz} \mbox{   , }
\end{eqnarray}
which is a simple approximation to the helioseismically determined 
profile (Schou {\it et al.} 1998; Gough, 2007). Here, $\Omega_{\rm eq}$ is the
observed equatorial rotation rate. As in Paper I, we 
finally select $\Omega_\odot$ to be 
\begin{equation}
\Omega_\odot = \Omega_{\rm eq} \left( 1 - \frac{a_2}{5} - \frac{3a_4}{35} \right)\mbox{   , }
\end{equation}
to ensure that the system has the same specific 
angular momentum as that of the imposed profile $\bu_{\rm cz}(r,\theta)$.

The computational domain is a spherical shell with the outer boundary 
located at $r_{\rm out} = 0.9 R_\odot$. It is chosen to be well-below the solar 
surface to avoid complications related to the very rapidly changing 
background in the region $r > 0.95 R_\odot$. The position of the lower 
boundary will be varied. 

The upper and lower boundaries are assumed to be impermeable. The
upper boundary is always stress-free, while the lower boundary is assumed to be 
either no-slip or stress-free depending on the calculation. 
In the no-slip case, the rotation 
rate of the excluded core is an eigenvalue of the problem, calculated 
in such a way as to guarantee that the total torque applied to the core is zero. 
Finally, the boundary conditions on temperature are selected in 
such a way as to guarantee that $\nabla^2 \tilde T = 0$ outside 
of the computational domain, as in Garaud \& Garaud (2008). 
We verified that the selection of the 
temperature boundary conditions only has a qualitative influence on the 
results, and doesn't affect the scalings derived. 

The numerical method of solution is based on the expansion of the governing
equations onto the spherical coordinate system, followed
by their projection onto Chebishev polynomials
$T_n(\cos\theta)$, and finally, solution of the resulting ODE system in $r$ 
using a Newton-Raphson-Kantorovich algorithm. The typical solutions 
shown have 3000 meshpoints and 60-80 Fourier modes. For more detail, see 
Garaud (2001) and Garaud \& Garaud (2008).

\subsection{The weakly stratified case}

We first consider the artificial limit of weak stratification. 
In the following numerical experiment, we use the available 
solar model background state, but divide the buoyancy frequency  
$\bar N$ by $10^3$ everywhere in the computational domain (all other 
background quantities remain unchanged). As a result, the new value of 
$\lambda_2$ in the domain is artificially reduced from the one 
presented in Figure \ref{fig:lambda2} by $10^3$, 
and is everywhere much smaller than one.
The position of the lower boundary is arbitrarily chosen to be 
at $r_{\rm in} =0.35 R_\odot$.  

\begin{figure}[h]
\centerline{\epsfig{file=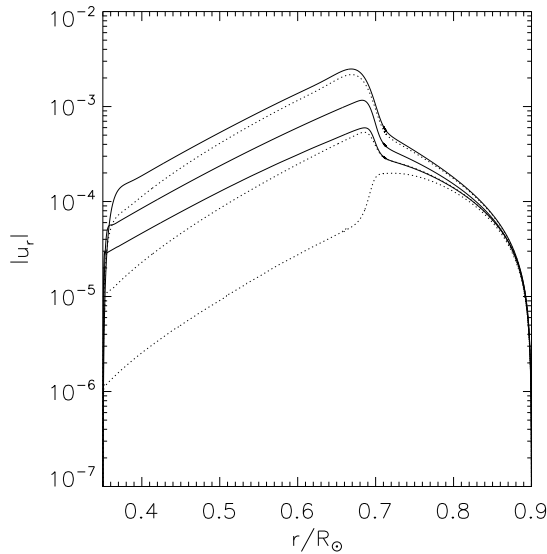,width=8cm}}
\caption{Vertical velocity at $80^{\circ}$ latitude in units of $R_\odot 
\Omega_\odot$ for an artificially 
weakly stratified simulation (where $\bar N$ 
was uniformly divided by $10^3$ everywhere).
The solid lines show three simulations for $f=10^{11}$,  $f=10^{10}$ 
and  $f=10^{9}$ (from top to bottom) for the no-slip case. These 
correspond to $\Enu = 2 \times 10^{-4}$,  $\Enu = 2 \times 10^{-5}$ 
and  $\Enu = 2 \times 10^{-6}$ at the base of the convection zone 
respectively, hence showing how $u_r \propto \Enu^{1/2}$. The dotted lines show simulations with 
stress-free boundary conditions for the same parameters, showing $u_r \propto \Enu$. In this 
calculation the overshoot  
depth $\Delta$ was selected to be 0.01$R_\odot$, and $\Lambda = 10$. 
The value of $D_0$ is irrelevant in this very weakly stratified simulation.}
 \label{fig:spher_ur_unstrat} 
\end{figure}

Two sets of solutions are computed for no-slip lower
boundary and for stress-free lower boundary. Figure \ref{fig:spher_ur_unstrat} 
is equivalent to Figure \ref{fig:unstr_bc2_wplot1}: it displays
the radial velocity $u_r$ as a function of radius near the poles 
(latitude of 80$^{\circ}$) for various values of $f$ -- in other words, 
$\Enu$ -- and clearly 
illustrates the scalings of $u_r \propto E_\nu^{1/2}$ in the radiative zone
for the no-slip case, and $u_r \propto E_\nu$ for the stress-free case.

\begin{figure}[h]
\centerline{\epsfig{file=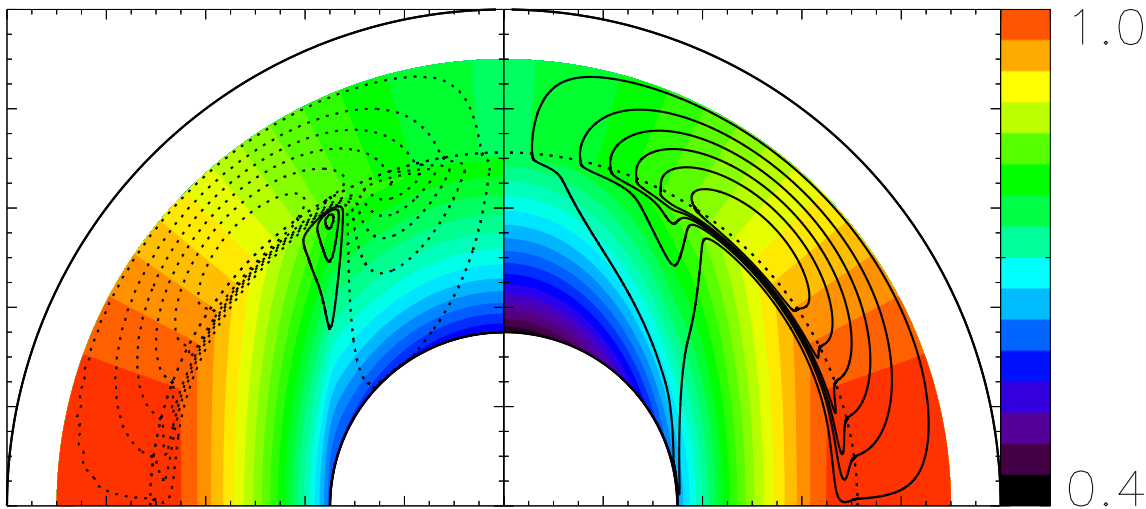,width=\textwidth}}
\caption{Normalized angular velocity ($\tilde{\Omega}/\Omega_{\rm eq}$) 
and streamlines solutions to equations 
(\ref{eq:global}) for an artificially weak stratification (see Figure \ref{fig:spher_ur_unstrat}), and for $f=10^9$ (corresponding 
to $\Enu = 2 \times 10^{-6}$ at the base of the convection zone). On the 
left, we show the solution with no-slip lower boundary conditions, 
and on the right the stress-free 
solution. Dotted lines represent clockwise flows, and solid lines 
counter-clockwise flows. In this calculation, the overshoot layer
depth $\Delta$ was selected to be 0.01$R_\odot$, and $\Lambda = 10$. 
The value of $D_0$ is irrelevant in this very weakly stratified simulation.}
 \label{fig:2D_spher_unstrat} 
\end{figure}

Figure \ref{fig:2D_spher_unstrat} illustrates the geometry of the flow 
in both no-slip and stress-free cases for $f = 10^9$ (which 
corresponds to an Ekman number near the radiative--convective interface of 
about $2 \times 10^{-6}$).  
The geometrical pattern of the flows observed within the convection zone
show a single-cell, with poleward flows near the surface and equatorward
flows near the bottom of the convection zone. Below the convection zone
we note the presence of three distinct regions: the polar region, 
a Stewardson layer region (at the tangent cylinder) and an equatorial 
region. Flows within the equatorial region are weak regardless of 
the lower boundary conditions. In the stress-free case, even in the 
tangent cylinder the flows tend to return mostly 
within the convection zone. If the lower boundary is no-slip on the
other hand, flows within the tangent cylinder are stronger, although the 
effect is not as obvious as in the Cartesian case because of the anelastic
mass conservation equation used here. 

\subsection{The stratified case}
\label{sec:stratspher}

Let us now consider the case of a true solar stratification. 
Since $\lambda_2$ increases rapidly with depth beneath the convection zone 
(from 0 to about 10 in the case of the Sun), the radius at which 
$\lambda_2 \simeq 1$ ($r_1$ for short) plays a special role: 
we expect the dynamics 
of the system to depend on the position of the lower boundary 
$r_{\rm in}$ compared with $r_1 \simeq 0.55R_\odot$. 

\begin{figure}[h]
\centerline{\epsfig{file=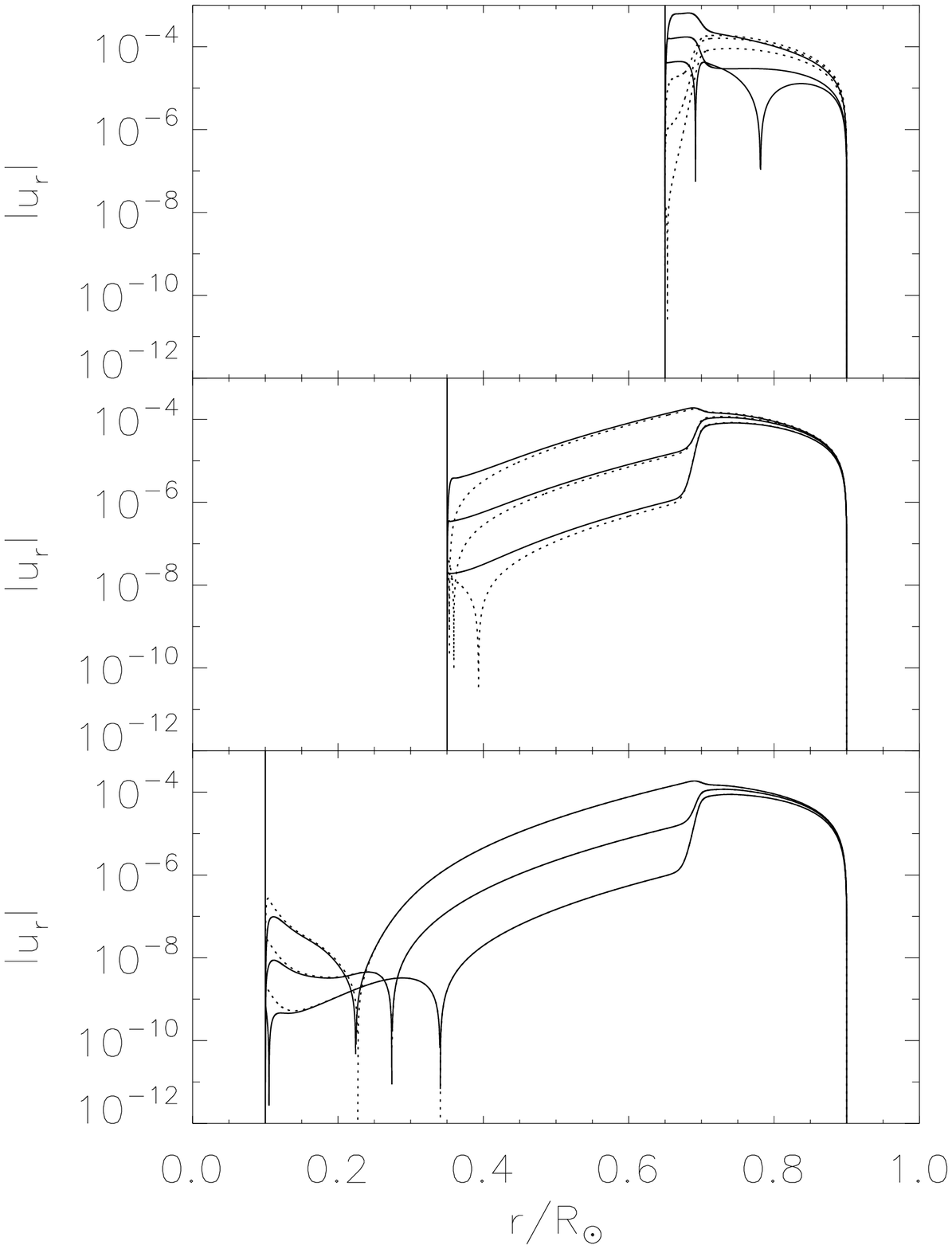,width=8cm}}
\caption{Vertical velocity (in units of $R_\odot \Omega_\odot$) 
in nine different simulations, at latitude $80^\circ$. The background
stratification in each case is solar, but the position of the lower boundary
is moved through the radiative zone from $0.65 R_\odot$ to $0.35 R_\odot$ and $0.1 R_\odot$. The solid-line plots are for no-slip lower boundary conditions while the dotted lines are for stress-free lower boundary conditions. Three simulations are shown in each case: (from lowest to highest curve) for $f=10^8$, $f=10^9$ and $f=10^{10}$ corresponding to $\Enu = 2\times 10^{-7}$ to $\Enu = 2\times 10^{-5}$. The logarithmic scale clearly shows that $u_r$ scales with $\Enu$ in the radiative zone in the stress-free cases for all values of $r_{\rm in}$ while in the no-slip case, $r_{\rm in}$ scales with $\Enu$ if $r_{\rm in} < 0.6$, as expected from Figure \ref{fig:lambda2}.} 
\label{fig:urrin} 
\end{figure}

This is 
indeed observed in the simulations, as shown in Figure \ref{fig:urrin}. If
$r_{\rm in} > r_1$, then $\lambda_2 < 1$ everywhere in the modeled section 
of the radiative zone. 
In this case, the dynamics follow the scaling for the unstratified case, 
and depend on the nature of the lower boundary ($u_r \propto \Enu$ if
the lower boundary is stress-free, and $u_r \propto \Enu^{1/2}$ if the 
lower boundary is no-slip). On the other hand, if $r_{\rm in} < r_1$
then the flows are strongly quenched 
by the stratification before they reach the lower boundary. As a result, 
the radial velocities scale with $\Enu/\lambda_2$ regardless
of the applied boundary conditions.

The implications of this final result, namely the importance of the 
{\it location} of the stresses involved in breaking the Taylor-Proudman 
constraint in relation to the radius at 
which $\lambda_2 \simeq 1$, are discussed in Section \ref{sec:applsun}.

\section{Implications of this work for solar and stellar mixing}
\label{sec:disc}

\subsection{Context for stellar mixing}
\label{sec:context}

The presence of mixing in stellar radiative zones has long been inferred 
from remaining discrepancies between models-without-mixing and
observations (see Pinsonneault 1997 for a review). The most 
commonly used additional mixing source is convective overshoot, 
whereby strong convective plumes travel beyond the radiative--convective
interface and cause intense but very localised (both in time and space)
mixing events (Brummell, Clune \& Toomre, 2002). 
The typical depth of the layer thus mixed, 
the ``overshoot layer'', is assumed to be a small fraction of a 
pressure scaleheight in most stellar models. 

A related phenomenon is wave-induced mixing (Schatzman, 1996). While most of
the energy of the impact of a convective plume hitting the stably 
stratified fluid below is converted into local buoyancy mixing, a fraction 
goes into the excitation of a spectrum of gravity waves, 
which may then propagate much further into the radiative interior. 
Where and when the waves eventually cause mixing 
(either through mutual interactions, thermal dissipation or by transferring 
momentum to the large-scale flow) depends 
on a variety of factors. It has recently been argued that the interaction
of the gravity waves with the local azimuthal velocity field 
(the differential rotation) would dominate the mixing process 
(Charbonnel \& Talon 2005), although this statement is not valid 
unless the wave-spectrum is near-monochromatic.
For the typically flatter wave spectra self-consistently 
generated by convection, wave-induced mixing has much more turbulent
characteristics (Rogers, MacGregor \& Glatzmaier 2008), and
is again fairly localized below the convection zone. 

Mixing induced by large-scale flows comes in two 
forms: turbulent mixing resulting from instabilities of the 
large-scale flows, and direct 
transport by the large-scale flows themselves. The former case is the dominant 
mechanism in the early stages of stellar evolution when the
star is undergoing rapid internal angular-momentum ``reshuffling'' 
caused by external angular-momentum extraction (disk-locked and/or jet phase, 
early magnetic breaking phase). In these situations, regions
of strong radial angular-velocity shear develop, which then become unstable
and cause local turbulent mixing of both chemical species and angular momentum. 
Studies of these processes were initiated
by the work of Endal \& Sofia (1978). Later, Chaboyer \& Zahn (1992) 
refined the analysis to consider the effect of stratification on 
the turbulence, and showed how this can differentially affect chemical 
mixing and angular-momentum mixing. Finally, Zahn (1992) proposed the 
first formalism which combines mixing by large-scale flows and mixing by 
(two-dimensional) turbulence. In addition to the flows driven by 
angular-momentum redistribution, he also considered flows driven by the local
baroclinicity of the rotating star, and showed that their effect 
can be represented as a hyperdiffusion term in the angular velocity
evolution equation. In a quasi-steady, uniformly rotating limit, the flows
described are akin to a local Eddington-Sweet circulation. His formalism is
used today in stellar evolution models with rotation (Maeder \& Meynet, 2000).

In all cases described above, mixing is either localised
near the base of the convection zone (overshoot, gravity-wave mixing), 
or significant only in very rapidly rotating stars (local Eddington-Sweet
circulations) or stars which are undergoing major angular-momentum 
redistribution (during phases of gravitational contraction, 
spin-down, mass loss, etc..).
In this paper, we have identified another potential cause of mixing, 
where the original energy source is the differential rotation in the 
stellar convective region: gyroscopic pumping (induced by the 
Coriolis force associated with the differential rotation, see McIntyre 2007) 
drives large-scale meridional flows which may -- under the right circumstances
-- penetrate the radiative region and cause a global circulation. 

This source of mixing is intrinsically {\it non-local} to the radiative zone. 
The simplest way of seeing this is to consider a thought-experiment where
the radiative--convective interface is impermeable: as shown
in Paper I, the amplitude of the meridional flows generated locally 
(i.e. below the interface) is then much smaller than the one calculated here.
The origin of the flows is also clearly independent of the baroclinicity 
(since the same phenomenon is observed in the unstratified limit), 
although the flows themselves can be influenced by the stratification. This
implies that they are not related to Eddington-Sweet flows. 
Finally, contrary to some of the other mixing sources listed above,
the one described here does not rely on the system being out-of-equilibrium:
it is an inherently quasi-steady phenomenon, implying that this process is an 
ideal candidate for ``deep mixing'' for stars on the Main Sequence.

The process together with the conditions under which strong mixing 
might occur, are now summarized and discussed.

\subsection{Qualitative summary of our results}

The differential rotation observed in stellar convective envelopes
(e.g. Barnes {\it et al.} 2005) is thought to be 
maintained by anisotropic Reynolds 
stresses, arising from rotationally constrained convective eddies 
(Kippenhahn, 1963).
The details of this particular process are beyond the scope of this 
study, but are the subject of current investigation by others 
(Kitchatinov \& R\"udiger, 1993, 2005; Rempel, 2005).
Instead, we have assumed here the simplest possible type of forcing 
which {\it mimics} the effect of convective Reynolds stresses in driving the 
system towards a differentially rotating state. Using this model, we 
then derive the expected mixing caused by large-scale meridional flows\footnote{
It is worth noting here that while we expect the details of the flow
structure and amplitude to be different when a more realistic forcing 
mechanism is taken into account, the overall scalings derived should not 
be affected. }.

We first found that large-scale flows are indeed self-consistently
driven by gyroscopic pumping in the convection zone, as expected 
(McIntyre, 2007). The amplitude of these flows {\it within the 
convection zone} scales roughly as 
\begin{equation}
V_{\rm cz} \sim \tau R_\star (\Delta \Omega) \mbox{   , }
\end{equation}
where $(\Delta \Omega)$ is the observed equator-to-pole 
differential rotation, $R_\star$ is the stellar radius, 
and $\tau$ is, as discussed in Section \ref{sec:modeleq}, 
related to the ratio of the convective turnover time divided by the 
rotation period. Note that for the Sun, with 
$(\Delta \Omega) \sim 0.1 \Omega_\odot$, the typical amplitude
of the corresponding meridional flows would be of the order of 
200 $\tau$ m/s -- which doesn't seem too unreasonable given the observations
of subsurface flows (Giles {\it et al.} 1997) 
and the typical values of $\tau $ in the solar convection zone 
(see Section \ref{sec:modeleq}).
 
Next, we studied how much mixing these flows might induce in the underlying
radiative zone. {\it In this quasi-steady formalism}, we found that the
magnitude of convection-zone-driven flows decays exponentially with depth
below the radiative--convective interface on 
the lengthscale $l_2$, where $l_2 = R_\odot /{\rm Pr}^{1/2}{\rm Ro}_{\rm rz}$, 
as determined in Section \ref{sec:stratrad} 
(see also Gilman \& Miesch, 2004 and 
Garaud \& Brummell, 2008). This penetration corresponds (in the linear regime) 
to a so-called ``thermo-viscous'' mode. The limit $l_2 \ll R_\odot$ corresponds
to a strongly stratified limit, where the flow velocities are rapidly 
quenched beneath the convection zone. The limit  $l_2 \gg R_\odot$ 
corresponds to the weakly stratified case, where the thermo-viscous 
mode spans the whole radiative interior and the 
stratification has little effect on the flow. 
It is important to note that ``weakly stratified'' 
regions in this context can either correspond to regions
with weak temperature stratification (small $\bar N$), 
{\it or} in rapid rotation, {\it or} with small Prandtl number 
-- this distinction will be used later. 

The amplitude of the flows upon entering the radiative zone $V_{\rm rz}$, 
together with $l_2$, uniquely define the global circulation timescale
in the interior (roughly speaking, $l_2/V_{\rm rz}$). 
In the weakly stratified/rapidly rotating limit, 
we find that the fraction of the meridional mass flux pumped in the convection 
zone which is allowed to {\it enter} the radiative zone is strongly 
constrained by Taylor-Proudman's theorem. This theorem, which holds when
the pressure gradient\footnote{more precisely, the perturbation to the 
pressure gradient around hydrostatic equilibrium} and the Coriolis force 
are the two dominant forces and are therefore in balance, 
enforces the invariance of all components of the flow velocities along 
the rotation axis. Hence, flows which enter the radiative zone 
cannot {\it return} to the convection zone unless the Taylor-Proudman constraint is broken. However, additional stresses (such as Reynolds stresses, viscous 
stresses, magnetic stresses) are needed to break this constraint.
As a result, two regimes may exist. If the (weakly stratified/rapidly rotating) 
radiative zone is in {\it pure} Taylor-Proudman balance, then the 
system adjusts itself, by adjusting the pressure field, in such a way as to 
ensure that the convection zone flows remain entirely within the convection 
zone. On the other hand, if there are other sources of 
stresses somewhere in the radiative zone to break
the Taylor-Proudman balance, then significant large-scale mixing is possible since
flows entering the radiative zone are allowed to return to the convection zone.
Furthermore, the resulting meridional mass flux in the radiative zone 
depends rather sensitively on the nature of the mechanism 
which breaks the Taylor-Proudman constraint (see next section). 

The strongly stratified/slowly rotating limit exhibits 
a very different behavior. Because of the strong buoyancy force,  
the Taylor-Proudman balance becomes irrelevant, the flows are 
exponentially suppressed, and the induced radiative zone 
mixing is independent of the lower boundary conditions. 
However, note that since $\bar N$ tends to 0 at a radiative--convective 
interface, there will always be a ``weakly stratified'' region in the 
vicinity of any convective zone. In that region the dynamics described
in the previous paragraph apply.

\subsection{Applications to the Sun and other stars}
\label{sec:applsun}

In the illustrative model studied here, the only stresses available
to break the Taylor-Proudman constraint are viscous stresses, which are 
only significant within the thin Ekman layer located near an artificial  
impermeable inner boundary. 
We do not advocate that this is a particularly relevant mechanism for the Sun! 
However, it is a useful example of the sensitive dependence of the 
global circulation mass flux on the mechanism responsible for breaking the 
Taylor-Proudman constraint. 

In the limit of weak stratification, we found that if the 
inner boundary is a stress-free boundary then the global turnover time
within the radiative zone is the viscous timescale. This is because stress-free
boundary conditions effectively suppress the Ekman layer. On the other hand
if the boundary layer is no-slip, then {\it the global mass flux through the 
radiative zone is equal to the mass flux allowed to return through the Ekman 
layer}. In that case, and according to well-known Ekman layer dynamics, the 
overall turnover time within the bulk of the domain is the geometric
mean of the viscous timescale and the rotation timescale 
($1/\Enu^{1/2}\Omega_{\odot}$), which correspond to a few million years only.

Going beyond simple Ekman dynamics, a much more plausible related scenario 
for the solar interior was studied by Gough \& McIntyre (1998).
They considered the same mechanism for the generation of large-scale flows 
within the convection zone, studied how these flows down-well into the 
radiative zone and interact with an embedded large-scale primordial magnetic 
field. They showed that the field can prevent the flows
from penetrating too deeply into the radiative zone, while the flows
confine the field within the interior. In their model, this 
nonlinear interaction occurs in a thin thermo-magnetic diffusion layer, 
located somewhat below the radiative--convective interface. 
One can therefore see an emerging 
analogy with the dynamics discussed here: in the Gough \& McIntyre model, 
the field does act as a somewhat impermeable barrier, 
and provides an efficient and elegant mechanism for breaking 
the Taylor-Proudman constraint within the
radiative zone. The only significant difference is that the artificial
Ekman layer is replaced by a more convincing thermo-magnetic 
diffusion layer: the mass flux allowed to down-well into the 
radiative zone, and mix its upper regions, is now controlled by 
a balance between the Coriolis force and magnetic stresses (instead of the 
viscous stresses). 
With this new balance, they find that the global turnover time 
for the circulation in the region between the base of the convection zone 
and the thermo-magnetic diffusion layer 
is of the order of a few tens of millions of years 
(which is still short compared with the nuclear evolution timescale).

This mixed region is the solar tachocline. By relating their 
model with observations, Gough \& McIntyre were able to identify the position 
of the magnetic diffusion layer to be just at the base of the observed 
tachocline (around $0.68 \pm 0.01 R_\odot$, see Charbonneau {\it et al.} 1999).
This turns out to be close enough to the base of the convection zone 
for the dynamics of the radiative region to be weakly-stratified in the sense
used in this paper (see Figure \ref{fig:lambda2} and Section 
\ref{sec:stratspher})
so that the meridional flows are indeed able to penetrate, and do so 
with ``significant'' amplitude (about $10^{-5}$ cm/s) down to the 
magnetic diffusion layer.
However, it is rather interesting to note that the Gough \& McIntyre model
could not have worked had today's tachocline been observed to be much 
thicker. It is also
interesting to note that for younger, more rapidly rotating solar-type stars, a
much larger region of the radiative zone can be considered 
``weakly stratified'', possibly leading to much deeper mixed regions if these
stars also host a large-scale primordial field. The implications of these
findings for Li burning, together with a few other interesting ideas, will 
be discussed in a future publication.

\section*{Acknowledgments}

This work was funded by NSF-AST-0607495, and the spherical domain simulations 
were performed on the UCSC Pleiades cluster purchased with an NSF-MRI grant. The authors thank N. Brummell  and S. Stellmach for many illuminating discussions, 
and D. Gough for the original idea.

\appendix

\section*{Appendix A: Ekman jump condition}

Equation (\ref{eq:unstr_radbc1_sol}) provides the solution ``far'' from the lower boundary, in the 
bulk of the fluid. Let us refer to the limit of bulk solutions 
as $z\rightarrow 0$ as $\hat u_{\rm bulk}(0^+)$ (and similarly for the other quantities). 
We now derive the Ekman solution close to the 
boundary, for the unstratified case. Let's study the problem using the stream-function $\psi$ with 
\begin{equation}
(\hat u,\hat v, \hat w) = \left(\hat u,\frac{\dd \hat \psi}{\dd z} ,-ik \hat \psi \right) \mbox{   .  }
\end{equation}
Moreover, let us assume that within the boundary layer, $\dd \hat \psi/\dd z \gg k \hat \psi$. The governing equations are then approximated by
\begin{eqnarray}
- 2 \frac{\dd \hat \psi}{\dd z} =  E_\nu \frac{ \dd^2 \hat u}{\dd z^2} \mbox{   ,  } \nonumber \\
2\hat u = - ik \hat p + E_\nu \frac{\dd^3 \hat \psi}{\dd z^3} \mbox{   ,  }  \nonumber \\
\frac{\dd \hat p}{\dd z} = - ik E_\nu \frac{\dd^2 \hat \psi}{\dd z^2} \mbox{   ,  } 
\end{eqnarray}
which simplify to
\begin{equation}
\frac{\dd^5 \hat \psi}{\dd z^5} = - 4 \frac{\dd \hat \psi}{\dd z}\mbox{   ,  } 
\end{equation}
with solutions 
\begin{eqnarray}
&& \hat \psi (z) = \psi_0 + \psi_1 e^{\lambda_3} +  \psi_2 e^{- \lambda_3 z} +  \psi_3 e^{\lambda_4 z} +  \psi_4 e^{-\lambda_4 z} \mbox{   ,  }  \nonumber \\
&& \hat u(z) = u_0 - \frac{2}{E_\nu} \left[ \frac{1}{\lambda_3} \psi_1 e^{\lambda_3 z} - \frac{1}{\lambda_3} \psi_2 e^{- \lambda_3 z} +   \frac{1}{\lambda_4} \psi_3 e^{\lambda_4 z} -  \frac{1}{\lambda_4}\psi_4 e^{-\lambda_4 z}   \right]\mbox{   ,  } 
\end{eqnarray}
where
\begin{equation}
\lambda_3 = (1+i) \Enu^{-1/2} \mbox{   ,  } \lambda_4 =(1-i) \Enu^{-1/2}\mbox{   .  } 
\end{equation}
The growing exponentials are ignored to match the solution far from the boundary layer; it then becomes clear that $u_0 = \hat u_{\rm bulk}(0^+)$, while $-ik \psi_0 = \hat w_{\rm bulk}(0^+)$. Requiring no-slip, impermeable conditions at $z=0$ implies
\begin{eqnarray}
\psi_0 + \psi_2 +  \psi_4 = 0 \mbox{   ,  } \nonumber  \\
\lambda_3 \psi_2  + \lambda_4 \psi_4 = 0\mbox{   ,  }  \nonumber  \\ 
u_0 - \frac{2}{E_\nu} \left[ -\frac{1}{\lambda_3} \psi_2 -   \frac{1}{\lambda_4} \psi_4 \right]  = 0\mbox{   ,  } 
\end{eqnarray}
which in turn implies 
\begin{eqnarray}
\psi_4 = - \frac{\lambda_3}{\lambda_4} \psi_2 \mbox{   ,  } \nonumber \\
\psi_2 = \frac{\lambda_4}{\lambda_3 - \lambda_4 }  \psi_0\mbox{   ,  }  \nonumber  \\ 
u_0 =  \frac{2}{E_\nu}  \frac{\lambda_3+\lambda_4 }{\lambda_3\lambda_4 } \psi_0
=  2 E_\nu^{-1/2} \psi_0\mbox{   .  } 
\end{eqnarray}
This last equation then uniquely relates the limit of the bulk solution 
$\hat u(0^+)$ and $\hat w(0^+)$ as $z\rightarrow 0$ as 
\begin{equation}
\hat u_{\rm bulk}(0^+) = \frac{ 2i}{k} E_\nu^{-1/2} \hat w_{\rm bulk}(0^+)\mbox{   ,  } 
\end{equation}
yielding the standard {\it Ekman jump condition}.
 
\section*{Appendix B: Stratified stress-free solution}

The boundary conditions discussed in Section \ref{sec:radstrat_noslip} imply the following set of equations. At $z=0$, $\hat w=0$ and $\hat u_z=0$ (alternatively, $\hat T = 0$):
\begin{eqnarray}
0 =  u_3  - u_4 \mbox{   ,  }  \nonumber \\
0 =  k u_1  - k u_2 + \lambda_2 u_3  - \lambda_2 u_4 \mbox{   .  } 
\end{eqnarray}
At $z=1$: $\hat w=0$ and $\hat T= 0$:
\begin{eqnarray}
0 = A e^{1/\delta} + B  e^{-1/\delta} - \frac{2iS}{k \Lambda} \mbox{   ,  } \\
0 = T_0 e^{k} + T_1 e^{-k}\mbox{   .  } 
\label{eq:appenB1}
\end{eqnarray}
Finally, matching conditions on $\hat w$, $\hat p$, $\hat T$ and $\dd \hat T/\dd z$ at $z=h$:
\begin{eqnarray}
- ik E_\nu \frac{k^2 - \lambda_2^2 }{\lambda_2} u_3  \sinh(\lambda_2 h) = A e^{h/\delta} + B  e^{-h/\delta} - \frac{2iS}{k \Lambda}\mbox{   ,  }  \nonumber \\
 2 u_1 \cosh(k h) + 2u_3 \cosh(\lambda_2 h) =   U_0(h)  + \frac{ik}{2} \delta \Lambda \left[ A e^{h/\delta} - B e^{-h/\delta} \right]\mbox{   ,  }  \nonumber \\
T_0 e^{kh} + T_1 e^{-kh} = -  \frac{4}{ik {\rm Ro}_{\rm rz}^{2}} \left[k u_1 \sinh(k h) +  \lambda_2 u_3 \sinh(\lambda_2 h) \right]\mbox{   ,  } \nonumber \\
T_0 e^{kh} - T_1 e^{-kh} = - \frac{4}{ik^2{\rm Ro}_{\rm rz}^{2}} \left[ k^2 u_1  \cosh(k h) +  \lambda_2^2  u_3 \cosh(\lambda_2 h) \right]\mbox{   ,  } 
\end{eqnarray}
where $u_4$ and $u_2$ were already eliminated using equations (\ref{eq:appenB1}). 
We now proceed to eliminate $A$, $B$, $T_0$ and $T_1$, which leaves two equations for $u_1$ and $u_3$:
\begin{eqnarray}
&& 2G\left[  u_1 \cosh(k h) + u_3 \cosh(\lambda_2 h)\right] - \delta \Lambda  k^2 E_\nu \frac{k^2 - \lambda_2^2 }{2 \lambda_2} u_3  \sinh(\lambda_2 h) \mbox{   ,  }  \nonumber \\
&& \quad =   \delta S \left( e^{(h-1)/\delta}(1 - G) -1 \right) +  G U_o(h) \mbox{   ,  }  \nonumber \\
&& \left( F - 1 \right) k u_1 \sinh(k h) +  k u_1  \cosh(k h) =  -  \frac{ \lambda_2^2}{k}  u_3 \cosh(\lambda_2 h) - \left( F - 1 \right)  \lambda_2 u_3 \sinh(\lambda_2 h)\mbox{   ,  } 
\end{eqnarray}
where the functions $F(h,k)$ and $G(h,h)$ are geometric factors defined as
\begin{eqnarray}
F(h,k) = \frac{2 }{1- e^{2k(h-1)}}\mbox{   ,  } \nonumber \\  
G(h,k) = \frac{ e^{(h-2)/\delta} - e^{-h/\delta}}{ e^{(h-2)/\delta} + e^{-h/\delta} }\mbox{   .  } 
\end{eqnarray}
These equations form a linear system for $u_1$ and $u_3$ with
\begin{eqnarray}
&& u_1 = - H u_3 \mbox{   ,  } \nonumber \\
&& u_3 = \frac{  \delta S \left( e^{(h-1)/\delta}(1 - G) -1 \right) +  G U_o(h) }{  2G\left[ -H \cosh(k h) + \cosh(\lambda_2 h)\right]  - \delta \Lambda  k^2 E_\nu \frac{k^2 - \lambda_2^2 }{2 \lambda_2}   \sinh(\lambda_2 h) } \mbox{   ,  } 
\end{eqnarray}
and where the function $H(h,k,\lambda_2)$ is given as 
\begin{equation}
H(h,k,\lambda_2) =   \frac{\lambda_2}{k} \frac{ \frac{ \lambda_2}{k}  \cosh(\lambda_2 h) + \left( F - 1 \right) \sinh(\lambda_2 h)}{      \cosh(k h)  + \left( F - 1 \right)  \sinh(k h) } \mbox{   .  } 
\end{equation}

These rather opaque solutions can be clarified a little by looking at the various relevant limits. For weakly stratified fluids
$\lambda_2 \rightarrow 0$. Then $H(h,k,\lambda_2) =  O(\lambda_2^2) \rightarrow 0$, and so
\begin{equation}
u_3 = \frac{  \delta S \left( e^{(h-1)/\delta}(1 - G) -1 \right) +  G U_o(h) }{  2G  - \delta \Lambda  k^4 E_\nu \frac{h}{2} } + O(\lambda^2_2)\mbox{   .  } 
\end{equation}
In the limit $\Enu \rightarrow 0$ this then becomes
\begin{eqnarray}
u_1 = O(\lambda_2^2) \mbox{   ,  } \nonumber \\
u_3 =\frac{1}{2} \left[  U_o(h) - \delta S  \frac{1 -  \cosh((1-h)/\delta)}{  \sinh((1-h)/\delta )} \right]\mbox{   .  } 
\end{eqnarray}
Folding this back into the original solution in the radiative zone then yields
\begin{equation}
\hat w(z) =  - ik^3 E_\nu   u_3  z = - \frac{ik^3 E_\nu }{2} \left[  U_o(h) - \delta S  \frac{1 -  \cosh((1-h)/\delta)}{  \sinh((1-h)/\delta )} \right] z\mbox{   ,  } 
\end{equation}
which is identical to equation (\ref{eq:unstrat_wrz}).

In the opposite, strongly stratified limit, $\lambda_2 \gg k$. Then we have instead 
\begin{equation}
H(h,k,\lambda_2) \simeq   \frac{\lambda_2^2}{k^2} \frac{\cosh(\lambda_2 h)}{\cosh(k h)  + \left( F - 1 \right)  \sinh(k h) } \mbox{   ,  } 
\end{equation}
so that this time $u_3 =  O(\lambda_2^{-2}) \rightarrow 0$, and in the limit $\Enu \rightarrow 0$ 
\begin{equation}
 u_1 =  \frac{1}{2\cosh(kh)} \left[  U_o(h) - \delta S  \frac{1 -  \cosh((1-h)/\delta)}{  \sinh((1-h)/\delta )} \right]\mbox{   .  } 
\end{equation}
Folding this back into the equation for $\hat w(z)$ in the radiative zone now yields
\begin{equation}
\hat w(z) = -  \frac{ik^3 \Enu }{2\lambda_2} \frac{\sinh(\lambda_2 z)}{\cosh(\lambda_2 h)} \left( 1 -\frac{\tanh(kh)}{\tanh(k(1-h))} \right) \left[  U_o(h) - \delta S  \frac{1 -  \cosh((1-h)/\delta)}{  \sinh((1-h)/\delta )} \right]\mbox{   ,  } 
\end{equation}
therefore justifying the scaling discussed in \ref{sec:match_strat_stressfree}.


\begin{thebibliography}{}
\bibitem{bal05}
 Barnes, J.R., Collier Cameron, A., Donati, J.-F., James, D. J., Marsden, S. C., \& Petit, P., 2005, MNRAS, 357, L1
\bibitem{bal02}
 Brummell, N. H., Clune, T. L., \& Toomre, J., 2002, \apj, 570, 825
\bibitem{cz92}
 Chaboyer, B. \& Zahn, J.-P., 1992, A\&A, 253, 173
\bibitem{c05}
 Charbonneau, P., 2005, LRSP, 2, 2
\bibitem{ct05}
 Charbonnel, C. \& Talon, S., 2005, Science, 309, 2189 
\bibitem{models}
 Christensen-Dalsgaard, J., {\it et al.}, 1996, Science, 272, 1286
\bibitem{eg99}
 Elliott, J. R. \& Gough, D. O., 1999, \apj, 516, 475
\bibitem{es78}
 Endal, S. \& Sofia, S., 1978, \apj, 229, 279
\bibitem{g01b}
 Garaud, P., 2001, PhD Thesis \\ available from http://www.ams.ucsc.edu/$\sim$pgaraud/
\bibitem{g07}
 Garaud, P., 2007. in {\it The Solar Tachocline}, pp. 147--181, eds. Hughes, D. W., Rosner, R. \& Weiss, CUP.
\bibitem{gb08}
 Garaud, P. \& Brummell, N. H., 2008, ApJ, 674, 498
\bibitem{gg08}
 Garaud, P. \& Garaud, J.-D., 2008, MNRAS, 391, 1239
\bibitem{gal97}
 Giles, P. M., Duvall, T. L., Jr., Scherrer, P. H. \& Bogart, R. S, 1997, Nature, 390, 52
\bibitem{gm04}
 Gilman, P. A. \& Miesch, M. S., 2004, \apj, 611, 568
\bibitem{gmi98}
 Gough, D. O. \& McIntyre, M. E., 1998, Nature, 394, 755
\bibitem{go07}
 Gough, D. O., 2007. in {\it The Solar Tachocline}, pp. 3--30, eds. Hughes, D. W., Rosner, R. \& Weiss, CUP.
\bibitem{k63}
 Kippenhahn, R., 1963, ApJ, 137, 664
\bibitem{kr93}
 Kitchatinov L. L. \& R\"udiger, G., 1993, A\&A, 276, 96  
\bibitem{kr05}
 Kitchatinov L. L. \& R\"udiger, G., 2005, Astr. Nachr., 326, 379
\bibitem{lh82}
 LaBonte, B. J. \& Howard, R. F., 1982, Solar Phys., 80, 361
\bibitem{mm00}
 Maeder, A. \& Meynet, G., 2000, ARA\&A, 38, 143
\bibitem{mci07}
 McIntyre, M. E., 2007. in {\it The Solar Tachocline}, pp. 183-212, eds. Hughes, D. W., Rosner, R. \& Weiss, CUP.
\bibitem{p97}
 Pinsonneault, M., 1997, ARA\&A, 35, 557
\bibitem{r05}
 Rempel, M., 2005, \apj, 622, 132
\bibitem{ral08}
 Rogers, T. M., MacGregor, K. B. \& Glatzmaier, G. A., 2008, MNRAS, 387, 616
\bibitem{r89}
 R\"udiger, G., 1989, in {\it Differential rotation and stellar convection. Sun and the solar stars}, Chapter 4. Publisher: Berlin, Akademie Verlag
\bibitem{s96}
 Schatzman, E., 1996, J. Fluid Mech., 322, 355
\bibitem{sal98}
 Schou, J., {\it et al.}, 1998, ApJ, 505, 390
\bibitem{sb68}
 Spiegel, E. A. \& Bretherton, F. P., 1968, ApJ, 153, L77
\bibitem{z92}
 Zahn, J.-P., 1992, A\&A, 265, 115
\end{thebibliography}
\end{document}